\documentclass[conference]{IEEEtran}
\IEEEoverridecommandlockouts
\usepackage{todonotes}
\usepackage{cite}
\usepackage{amsmath,amssymb,amsfonts}
\usepackage{textcomp}
\usepackage{wrapfig}
\usepackage{algorithmic}
\usepackage{graphicx}
\usepackage{textcomp}
\usepackage{xcolor, cancel}
\usepackage{array}
\usepackage{tabularx}
\usepackage{ctable}
\usepackage{multirow, multicol}
\usepackage{threeparttable}
\usepackage{color, colortbl}
\usepackage{caption,subcaption}
\usepackage{balance}
\usepackage{hyperref}
\usepackage{cleveref}
\usepackage{amsmath}
\usepackage[normalem]{ulem}
\crefformat{section}{\S#2#1#3}

\makeatletter
\renewcommand{\fnum@figure}{Figure. \thefigure}
\makeatother

\definecolor{Gray}{gray}{0.90}

\newcommand*\circled[1]{\tikz[baseline=(char.base)]{
            \node[shape=circle,draw,inner sep=0.8pt] (char) {#1};}}

\newcommand{\projectname}{\texttt{ARcode}}

\def\BibTeX{{\rm B\kern-.05em{\sc i\kern-.025em b}\kern-.08em
    T\kern-.1667em\lower.7ex\hbox{E}\kern-.125emX}}
    
\colorlet{updates}{blue}

\begin{document}

% \title{Conference Paper Title*\\

\title{ARcode: HPC Application Recognition Through Image-encoded Monitoring Data}

\author{\IEEEauthorblockN{1\textsuperscript{st} Jie Li}
\IEEEauthorblockA{\textit{Department of Computer Science} \\
\textit{Texas Tech University}\\
Lubbock, TX, USA \\
jie.li@ttu.edu}
\and
\IEEEauthorblockN{2\textsuperscript{nd} Brandon Cook}
\IEEEauthorblockA{\textit{National Energy Research Scientific Computing Center} \\
\textit{Lawrence Berkeley National Laboratory}\\
Berkeley, CA, USA \\
bgcook@lbl.gov}
\and
\IEEEauthorblockN{3\textsuperscript{rd} Yong Chen}
\IEEEauthorblockA{\textit{Department of Computer Science} \\
\textit{Texas Tech University}\\
Lubbock, TX, USA \\
yong.chen@ttu.edu}
}
% \author{Anonymized for double-blind review}

\maketitle

\begin{abstract}

Knowing HPC applications of jobs and analyzing their performance behavior play important roles in system management and optimizations. 
% They pave the way to design optimal resource allocation strategies, achieve efficient energy efficiency, and prevent malware programs from stealing computing cycles. 
The existing approaches detect and identify HPC applications through machine learning models. However, these approaches rely heavily on the manually extracted features from resource utilization data to achieve high prediction accuracy.
% Having knowledge of applications running on HPC systems is one of the most important ways to design optimal resource allocation strategies, provide efficient energy management, and take preventive actions when facing anomalies. However, the state-of-the-art approach is to identify applications by resource utilization patterns, which involves extensive manual feature construction to achieve high prediction accuracy. 
In this study, we propose an innovative application recognition method, \projectname, which encodes job monitoring data into images and leverages the automatic feature learning capability of convolutional neural networks to detect and identify applications. Our extensive evaluations based on the dataset collected from a large-scale production HPC system show that \projectname\ outperforms the state-of-the-art methodology by up to 18.87\% in terms of accuracy at high confidence thresholds. For some specific applications (BerkeleyGW and e3sm), \projectname\ outperforms by over 20\% at a confidence threshold of 0.8.

\end{abstract}


\section{Introduction}\label{introduction}

As HPC systems are approaching the exaFLOP era, the scale and complexity of HPC systems have increased significantly over the past few years. Administrators need to understand not only the performance of the hardware system, but also the typical applications and their characteristics, such as how they use the computing resources and how they have been executed before~\cite{brandt2009resource, allcock2011challenges, jones2012application, evans2014comprehensive, li2020monster}. With the increase of computation capability, the resource contention and energy consumption increase as well. To improve HPC system efficiency, it is imperative to understand the characteristics of applications and to guide better resource-aware scheduling policies based on the knowledge of resource requirements of applications~\cite{shoukourian2014predicting, dutot2017towards, tuncer2018online}. Moreover, emergent misbehaviour is becoming more prevalent due to the large scale and high utilization~\cite{dean2009large}. For system administrators striving to guarantee optimal system performance, detecting anomalies and potential errors of applications is an essential task~\cite{baseman2016interpretable, dani2017k, borghesi2019online}.

% unauthorized applications such as bitcoin mining programs could take advantages of the high computing capability and consume computing hours that are for scientific discoveries. Therefore, knowing the applications that are running on HPC will help administrators ban these malware programs in a proactive manner~\cite{garcia2015anomaly, zhou2018hardware, kuruvila2020analyzing, gangwal2020detecting}. 

An online detection system that is capable of identifying applications in real-time, with little or no human intervention, would be a boon to system management. However, this is a daunting task. Large-scale HPC systems are generally shared by a variety of users from different domains. In addition to traditional large-scale simulation applications (e.g., molecular dynamics, quantum chemistry applications, and climate simulations), emerging Machine Learning (ML) and Artificial Intelligence (AI) applications have become an increasingly critical part of the workloads on HPC systems. Frequently used applications and libraries are usually pre-installed on the system by system administrators. However, administrators do not necessarily possess the knowledge about the executions and characteristics of each application. Users could build, compile and name their own applications that are not shared with others. In addition, if users do not provide application information in job submission scripts, it is difficult to know what applications these jobs belong to. These various use cases and the imprecise mapping of job names to actual applications make it difficult to identify applications using naive approaches. As an example, we examined the application names derived from the job submission script on Cori and found that about 42.4\% of the names were incorrect. 

% Such difficulties hinder the opportunity to develop optimal scheduling polices and to take appropriate action when faced with anomalies such as errors and unwanted behaviors.

% blocking malicious applications.

% Due to the privacy policy, users’ data and codes are not exposed to the system administrators. The only available information about the applications is the job submit scripts that are used for submitting jobs through job scheduler. These various of use cases as well as the privacy policy render the difficulties of analyzing the applications, and therefore hinder the opportunities for developing optimized scheduling polices and blocking malicious applications.

% However, the binary based approach cannot be applied in the HPC environment, where it is impractical to conduct binary differencing among hundreds of thousands of executables. In addition, obtaining and managing binaries of users’ applica- tions are not always possible for HPC researchers.

% no-ML approaches: metadata and binary signatures. 
% metadata: lacks of accuracy; 
% Binary signatures: binaries collection/ incorrect if compiled with different compiler toolchain

Advanced methods for detecting and recognizing applications can be divided into \textit{static analysis} of binaries and/or scripts, which can be performed without running a job, and \textit{dynamic analysis} of system logs and performance metrics, which implies analysis during or after job execution. Early works explored static analysis of binaries to determine the semantic similarity between two applications~\cite{flake2004structural, dullien2005graph}. However, The complexity of HPC systems has reached a point where static analysis of the binaries used to run and maintain the detection system is no longer feasible. This approach is invasive to users' data and requires a dedicated binaries collection and management infrastructure, which is not always of interest for system administrators. Moreover, even though the binaries are available, it does not perform well if the same application is compiled by a different compiler toolchain or optimization level~\cite{gao2008binhunt, bourquin2013binslayer,egele2014blanket}.

Collecting and analyzing system logs and performance metrics are critical to combat performance crisis, and they are prevalent in HPC systems, although the approach may vary. Recently, there has been a growing research interest in automatic detection that relies on extensive performance metrics and employs ML techniques to identify applications~\cite{combs2014power, ates2018taxonomist, zou2019fingerprinting, ramos2019accurate, jakobsche2021execution}. A representative approach proposed by Ates et al. explored building supervised ML models with  statistic features of monitoring metrics to classify applications~\cite{ates2018taxonomist}. The classification model relies on thousands of statistical features extracted from hundreds of time-series monitoring metrics to achieve high prediction accuracy. A major weakness of this approach is the high-latency responses of the detection model, and the statistical features are only accurate for representing the application after the job is finished. In addition, the performance of feature-based models is highly dependent on feature engineering; using different features has the potential to deviate the classification performance~\cite{zhao2017convolutional}.

% Feature engineering is the process of using domain knowledge of the data to create features that make machine learning algorithms work. 

In this study, we extend the line of performance metrics based approaches and propose an innovative method called \emph{\projectname} (stands for \textbf{A}pplication \textbf{R}ecognition \textbf{code}). It is an application recognition method utilizing images encoded from performance monitoring metrics. Specifically, we leverage monitoring metrics collected from HPC systems (\cref{cori}) and encode time-series data to two-dimensional images to represent the resource usage patterns of HPC executions (or \emph{job signatures} for simplicity, discussed in \cref{signature}). We then build a dataset labelled with application names and train a Convolutional Neural Network (CNN) to build the classification model (\cref{model}). The contributions of this study are summarized below: %The encoded image, which we named \emph{job signature}, is a representation of monitoring traces that preserve temporal performance behavior of the job.

\begin{itemize}

\item Contrary to other studies where datasets are generated from benchmarks and proxy applications, our dataset is built from real applications with different input data, resource allocations and run times, which well reflects the complex real scenarios. Specifically, we collect monitoring data from a production system and build a dataset of performance metrics of twelve popular HPC applications where the application names are labelled. 

% In high performance computing (HPC), proxy applications (“proxy apps”) are small, simplified codes that allow application developers to share important features of large applications without forcing collaborators to assimilate large and complex code bases.

\item Our innovative methodology encodes time-series monitoring data into two-dimensional images, where the performance metrics are creatively represented in a much smaller size compared to the original data without losing important metric variations. The encoded job signatures can be used not only for application classification and detection, but also to inspire methods for predicting and estimating the resource usage of applications.

\item We use the CNN techniques and train the CNN model with the job signatures. The job signatures are generated from the performance monitoring data, thus do not involve collecting and analyzing users' private data. The CNN model, on the other hand, does not require manual features engineering, making it easier to be tuned and adopted by any HPC sites.

\end{itemize}

Through extensive experiments, we find that \projectname\ achieved a competitive classification performance in most cases and outperformed by up to 18.87\% at high confidence thresholds compared to the state-of-the-art methods. When detecting some specific applications (e.g., BerkeleyGW and e3sm) with a confidence threshold of 0.8, \projectname\ is better than the state-of-the-art methods by over 20\% in terms of accuracy. Meanwhile, \projectname\ retains the temporal information of the monitoring data and is able to recognize running jobs. This capability is not available in any state-of-the-art methods. The details of all these experiment evaluations are discussed in ~\cref{experiments}. The \projectname\ model and dataset used in this study can be found in a separate submission of artifacts. 
\section{Background}\label{background}
In this section, we briefly introduce the Cori system and how we collect job monitoring data. Then, we describe the workflow of the monitoring infrastructure being used and present the available job-level monitoring metrics.

\subsection{The Cori System}\label{cori}

Cori\footnote{\url{https://docs.nersc.gov/systems/cori/}} is a Cray XC40 system at National Energy Research Scientific Computing Center (NERSC). It consists of 2,388 Intel Xeon ``Haswell'' processor nodes and 9,688 Intel Xeon Phi ``Knight's Landing'' (KNL) nodes interconnected on Cray Aries High-Speed Network, which provides a peak performance of about 30 petaflops. Additionally, Cori is equipped with a large scratch Luster File System that provides 432 GB/s of performance with a capacity of 28.5 petabytes to the compute nodes. Cori also has the Cray DataWarp based Burst Buffer, offering a 1.8 petabytes burst buffer storage with 1.7 TB/s in peak bandwidth performance~\cite{he2018preparing, antypas2014cori}. 
With the mission of accelerating the pace of scientific discovery through HPC and data analysis, workloads running on the Cori system cover a wide range of scientific disciplines, including lattice QCD, materials science, climate research, high energy physics, astrophysics, and more. 

\subsection{Monitoring Workflow}

\begin{figure}
\centering
\includegraphics[width=0.7\linewidth]{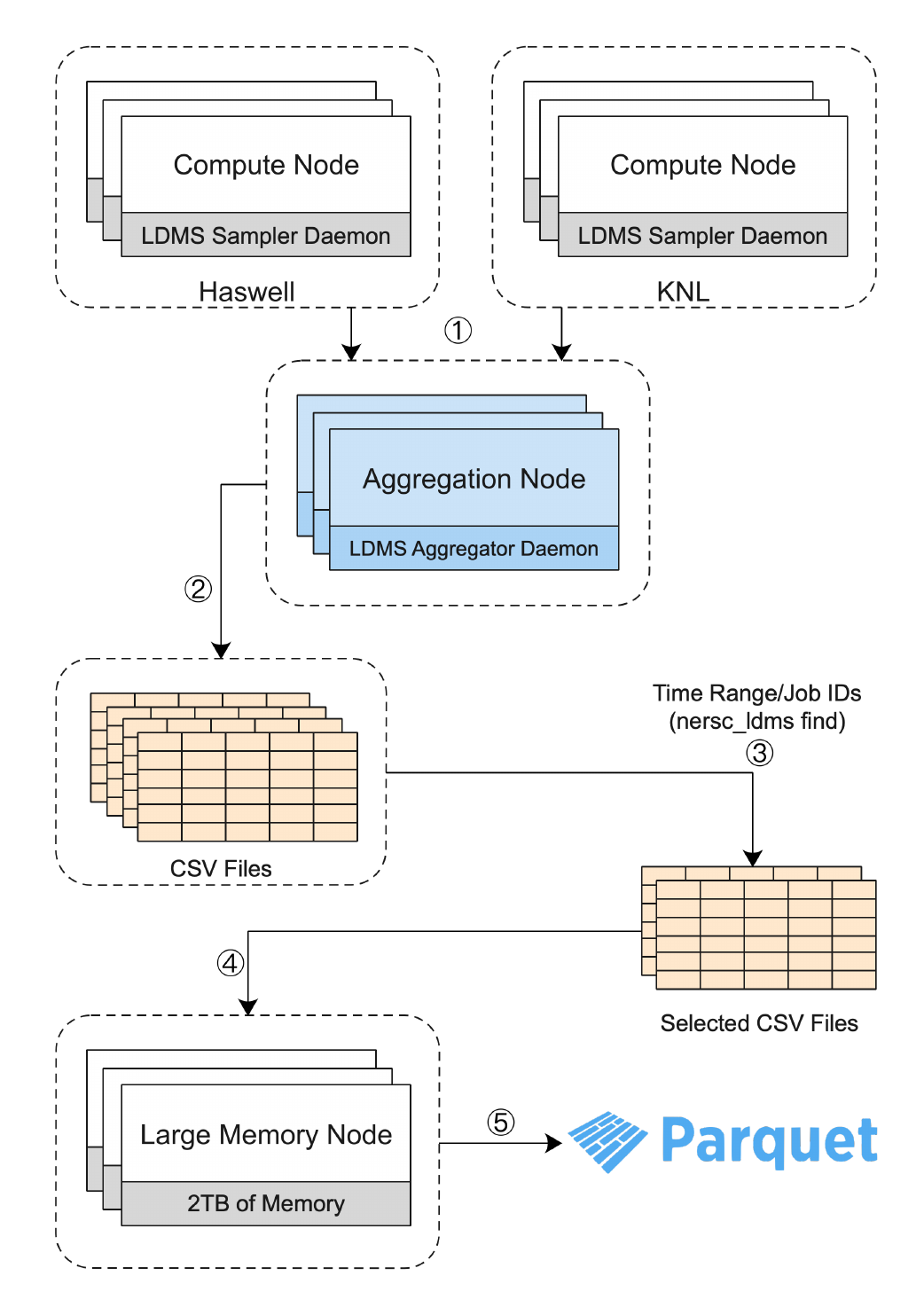}
\caption{Workflow of LDMS on Cori}
\label{fig:workflow}
\end{figure}

\begin{table*}
    \centering
    \caption{Selected Monitoring Metrics}
    \label{table:metrics}
    \begin{tabular}{ c | c | c | c }% {|c|c|c|}
        \specialrule{.1em}{.05em}{.05em}
        \rowcolor{Gray}
        \textbf{Sampler} & \textbf{Metrics} & \textbf{Derived Metrics} & \textbf{Description}\\
        \specialrule{.1em}{.05em}{.05em}
        cray\_aries\_sampler & power & power & Node Power Consumption\\
        \hline
        \multirow{2}{*}{syspapi} & \multicolumn{1}{c|}{PAPI\_TOT\_INS} & \multirow{2}{*}{IPC (PAPI\_TOT\_INS/PAPI\_TOT\_TOT)} & \multirow{2}{*}{Instruction Per Cycle}\\
        \cline{2-2}
        & \multicolumn{1}{c|}{PAPI\_TOT\_TOT} & & \\
        \hline
        \multirow{2}{*}{meminfo} & \multicolumn{1}{c|}{MemTotal} & \multirow{2}{*}{Mem (MemTotal - MemFree)}  & \multirow{2}{*}{Memory Used }\\
        \cline{2-2}
        & \multicolumn{1}{c|}{MemFree} & \\
        \hline
    \end{tabular}
\end{table*}

\begin{figure*}[t]
    \centering
    \includegraphics[width=0.85\linewidth]{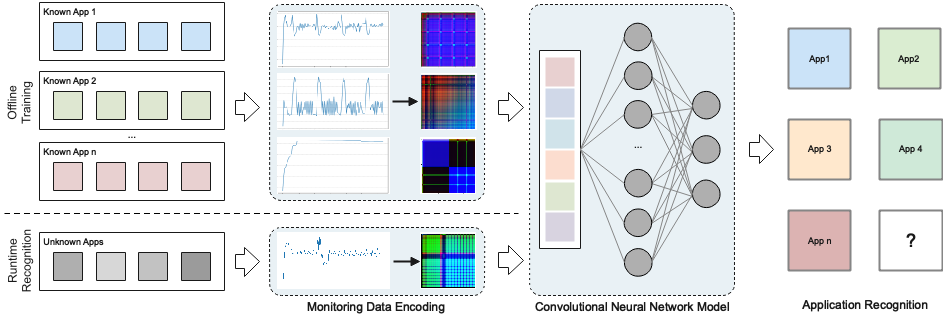}
    \caption{Overview of \projectname\ Design. \projectname\ encodes the time-series monitoring data into job signatures. In the offline training phase, \projectname\ trains the CNN from a labelled job signature dataset. In the running recognition phase, the CNN model is used to detect the encoded job signature. The CNN model is modified so that it can identify novel applications.}
    \label{fig:overview}
\end{figure*}

% The complex HPC systems require monitoring each components in the system to maintain the health of the system, provide stable computing services to HPC users, and carry out performance tuning. Based on the data source of metrics, there are two categories of monitoring: one is in-band monitoring, where the performance metrics are collected from operating system kernels and hardware performance counters; another is out-of-band monitoring, where dedicated sensors, such as temperature sensors, power meters, are attached to the compute nodes to collect performance metrics. 

In this study, we utilize the monitoring metrics collected from the CPU nodes of the NERSC Cori system, where the in-band monitoring tool, \emph{LDMS} is used \cite{agelastos2014lightweight}. 
% LDMS stands for Lightweight Distributed Metric Services, it is a low-overhead, low-latency framework for collecting, transferring, and storing metric data. 
The detailed discussion of how LDMS works on Cori is out of scope of this paper. In Figure~\ref{fig:workflow}, we illustrate its workflow for generating job-level performance metrics. The workflow includes the following steps: \circled{1} LDMS samplers on Haswell and KNL nodes (two partitions in Cori) collect in-band metrics at a pre-configured frequency; \circled{2} the monitoring data are then sent to aggregation nodes, where metrics collected from the same sampler are stored in CSV files under the same folder. Each CSV contains metrics from multiple nodes and the corresponding metrics for a job may span multiple CSV files. 
To improve the usability of the monitoring data, the \emph{NERSC\_LDMS}~\cite{nerscldms}, an LDMS data processing tool, takes care of the post-processing of CSV files. \circled{3} NERSC\_LDMS gets job IDs from Slurm \emph{sacct} and joins the job IDs with CSV files, and \circled{4} submits these files to large memory nodes for post-processing, where the same metrics of the same job are extracted. \circled{5} The post-processed job-level metrics are saved in parquet files by metric samplers for future analysis. Compared to raw CSV files, job-level parquet files significantly improve the availability of monitoring data and the efficiency of querying job performance metrics.

\subsection{Available Metrics}
The available metrics collected through LDMS on HPC systems depend on the configurations and available samplers on the hardware platform. As routine tasks of monitoring the health of the Cori system, 34 different samplers collect metrics related to I/O, network, CPU counters, memory usage, power consumption, etc. The total number of metrics collected through the sampler varies from 12 to 3,016, depending on the sampler. The granularity of collected metrics is one second, generating approximately 400MB of monitoring data per second on a system wide basis. 

From the perspective of application recognition, it is unattractive and impractical to include all of these extensive monitoring metrics in one model because processing large amounts of time-series data is compute-intensive and time-consuming. On the other hand, we envision that the proposed methodology should be easily adopted by other HPC systems and the selected metrics should be common even when using different monitoring infrastructures. Therefore, we select \emph{five} of these metrics and derive \emph{three} representative metrics for creating job signatures. These three metrics are the power consumption of the compute node, instruction per cycle (IPC), and memory used as shown in Table~\ref{table:metrics}. These metrics are expected to be available through the monitoring infrastructure on a variety of HPC architectures.

\section{Metrics Encoding and Classification}

In this section, we first discuss design considerations and provide an overview of \projectname\ design. We then present the details of encoding monitoring data, including resampling job metrics, converting 1D time-series data into 2D images, and encoding multiple traces into a single image. Lastly, we introduce the CNN architecture for classifying the encoded images. 

\subsection{Design Consideration and Overview}

\begin{figure*}[t]
     \centering
     \begin{subfigure}[b]{0.35\textwidth}
         \centering
        \includegraphics[width=0.95\linewidth]{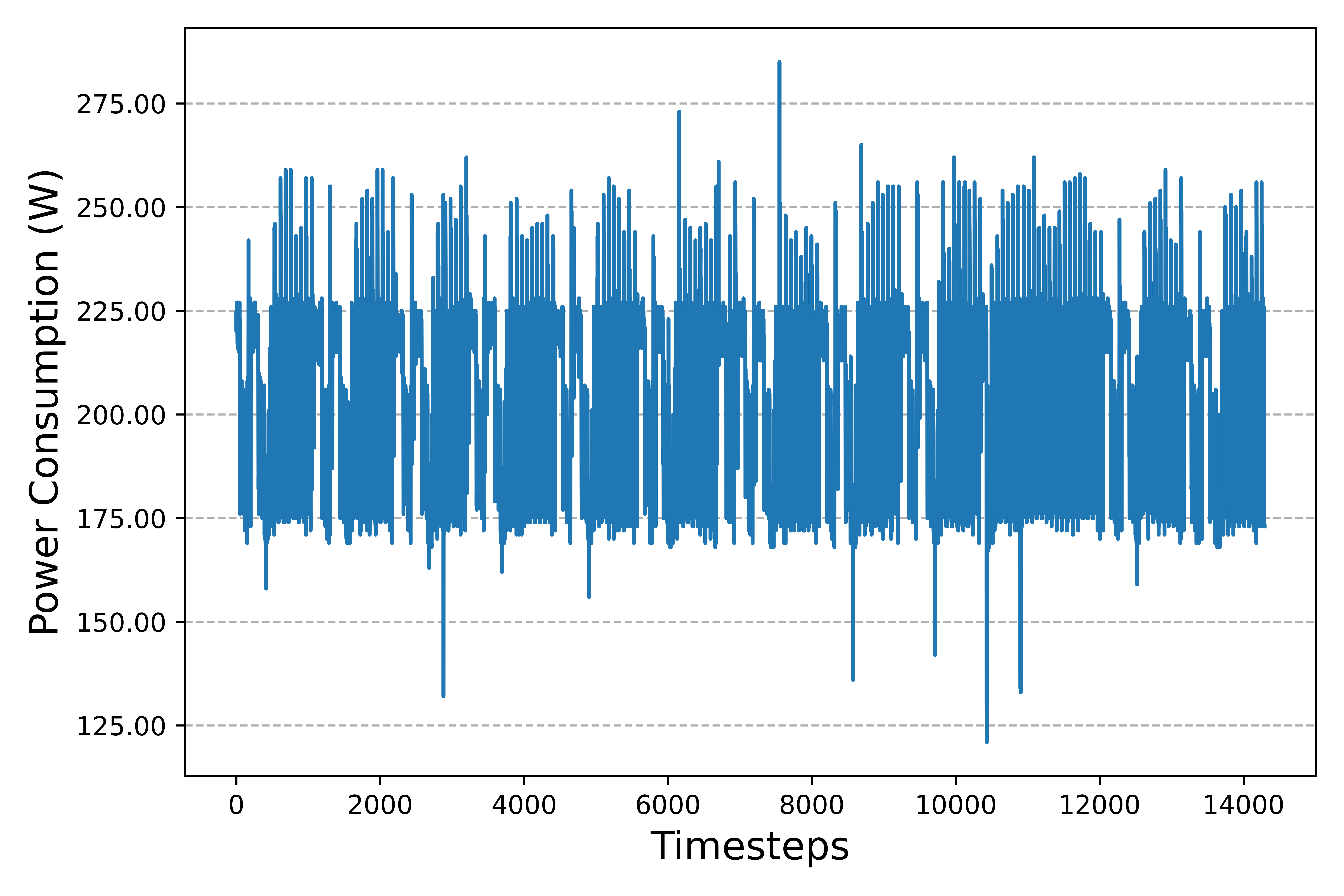}
        \caption{Original Power Consumption Trace}
     \label{fig:downsampling_raw}
     \end{subfigure}
     \hspace{2em}
     \begin{subfigure}[b]{0.35\textwidth}
         \centering
        \includegraphics[width=0.95\linewidth]{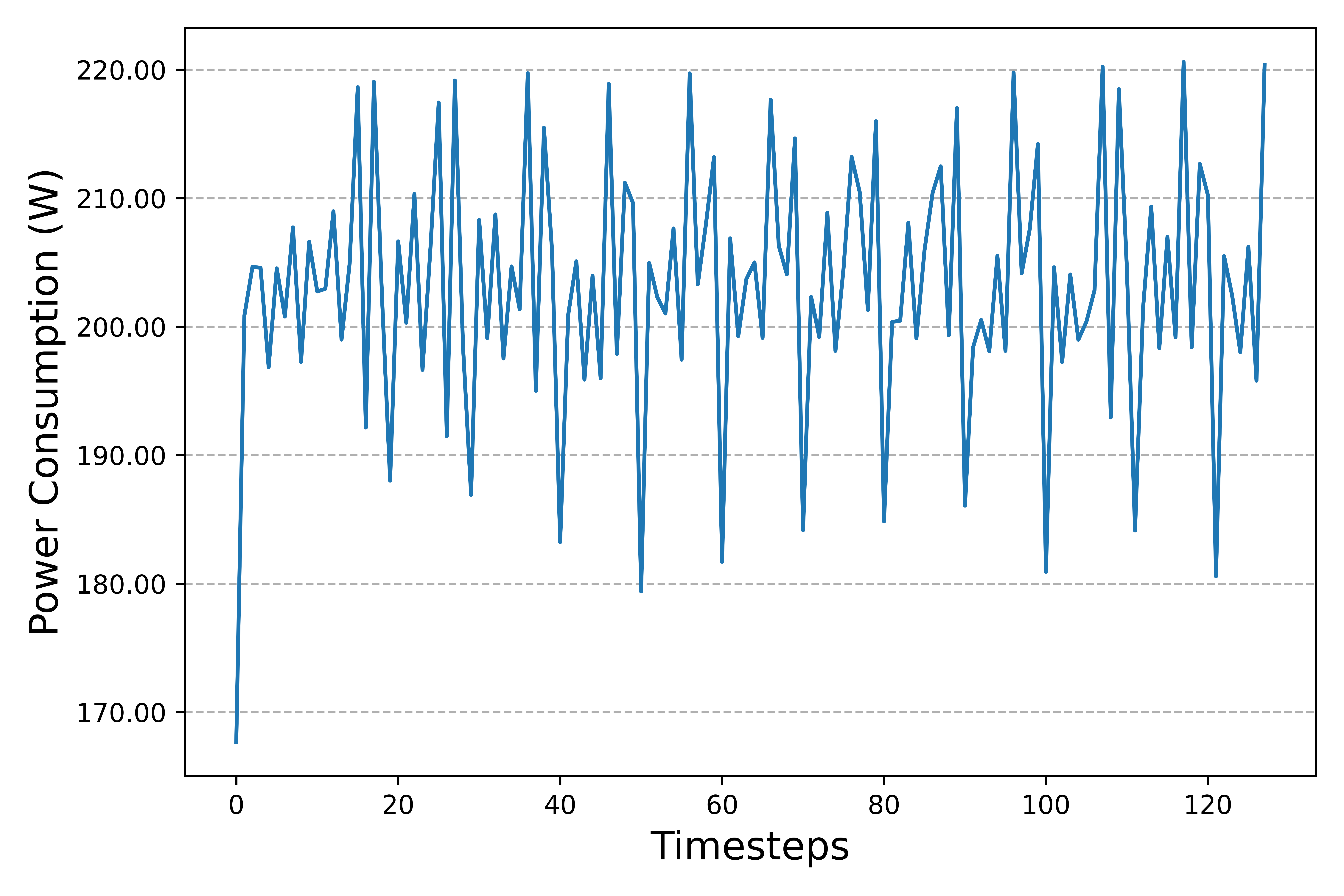}
        \caption{Downsampled Trace by Aggregating}
     \label{fig:downsampling}
     \end{subfigure}
     
     \centering
     \begin{subfigure}[b]{0.35\textwidth}
         \centering
        \includegraphics[width=0.95\linewidth]{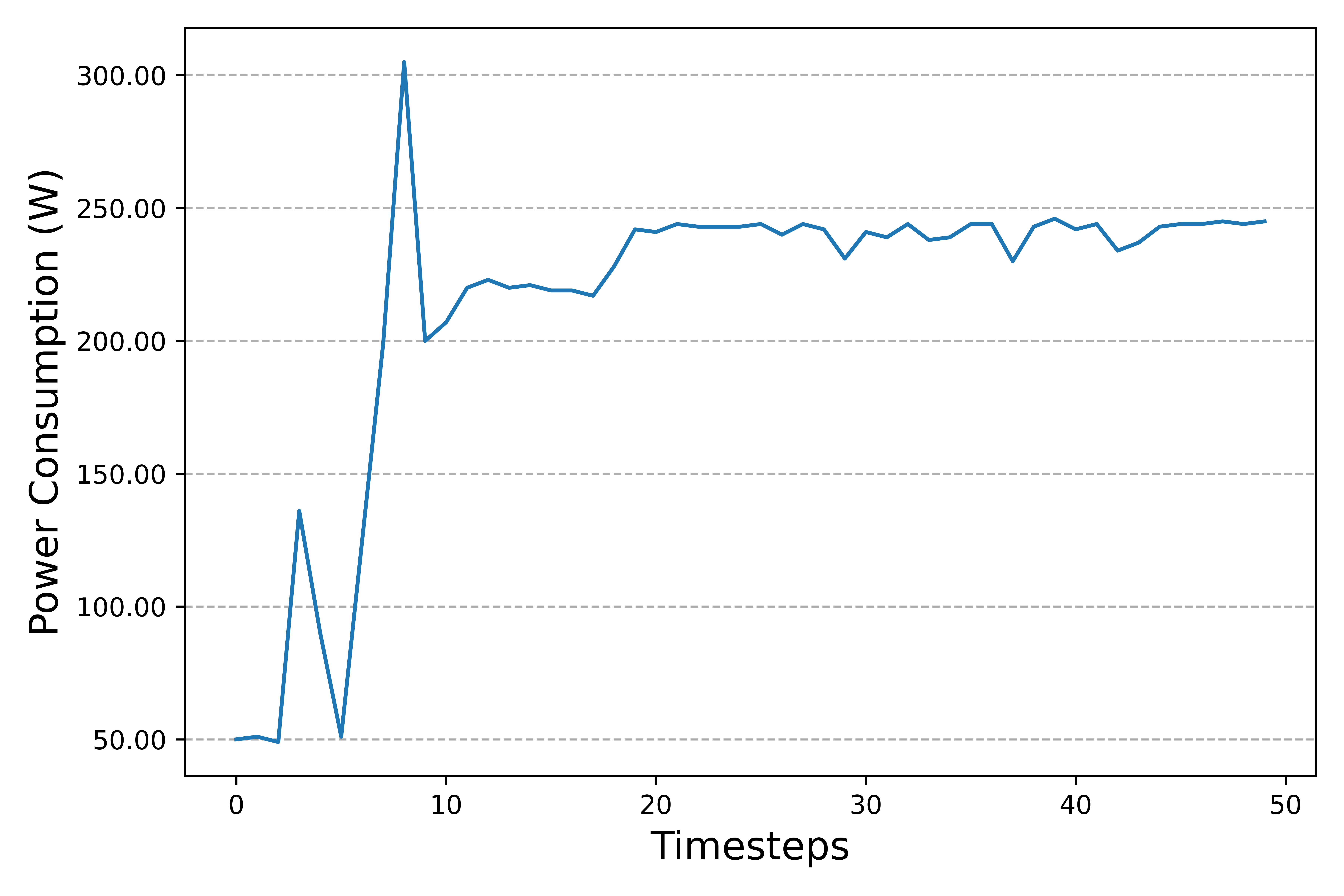}
        \caption{Original Power Consumption Trace}
     \label{fig:upsampling_raw}
     \end{subfigure}
     \hspace{2em}
     \begin{subfigure}[b]{0.35\textwidth}
         \centering
        \includegraphics[width=0.95\linewidth]{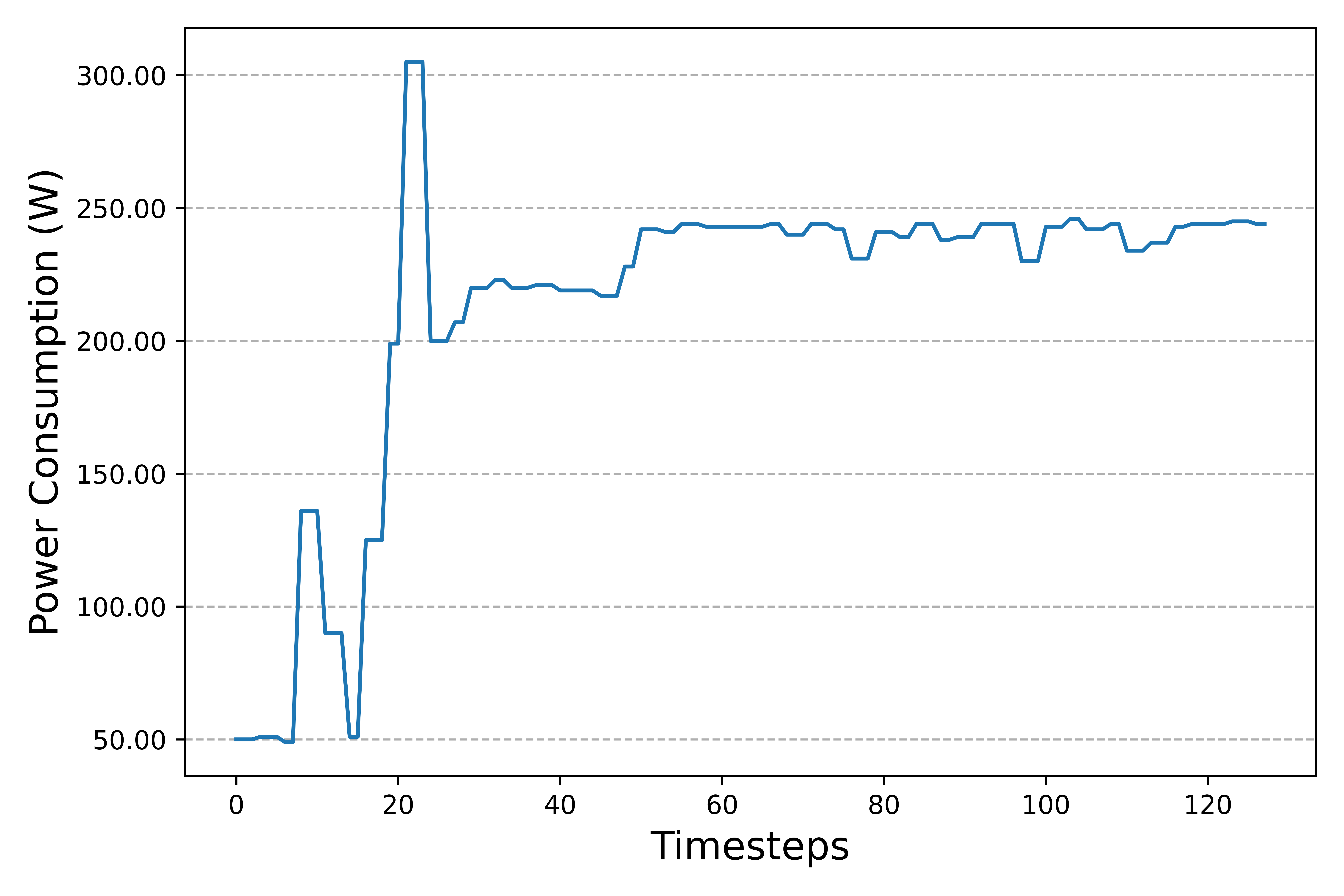}
        \caption{Upsampled Trace by Padding}
     \label{fig:upsampling}
     \end{subfigure}
     
    \caption{Resampling traces by aggregating and padding based on the original trace length and the predefined length. Figure a is the trace longer than the predefined length 128, in which the aggregating function is applied to downsample the trace. Figure c is a trace shorter than 128, in which the padding is used to upsample the trace. Figure b and d show the corresponding resampled traces, respectively.}
    \label{fig:resampling}
\end{figure*}

As discussed in~\cref{introduction}, existing solutions either perform a static analysis of binaries and/or scripts or leverage machine learning methods to extract key features out of extensive monitoring data to build application prediction and classification models. 
%
%Binaries and/or scripts based approaches require dedicated collection and management infrastructures, which are not as prevalent as monitoring infrastructures for performance metrics. 
%
Our approach, \projectname, is similar to the monitoring data based approaches but with two design considerations: features learned without human intervention and data retaining temporal information.

First, current state-of-art application detection models, such as Taxonomist~\cite{ates2018taxonomist}, extract statistical summaries from raw monitoring data to create a feature vector for machine learning models, where the classification performance is highly dependent on the quality of the manually constructed features. To improve the usability, one of our considerations is that the features of the monitoring traces can be learned without human intervention. Second, although statistical features are spatially efficient and lightweight when building detection models, they lose the temporal information of time-series data, thus limiting their use case. To improve the extensibility, the other consideration of our model is to retain temporal information. So it is able to use part of the monitoring data in detection, classification, and prediction of the resource usage of applications.

% --------------------------------------------

With these considerations in mind, our proposed approach, \projectname, encodes entire time-series monitoring data of jobs into unified-size images and leverages deep learning techniques to learn features. The encoded image, which we named as \textit{job signature}, is a representation of monitoring traces that preserve temporal performance behavior of the job. 

As shown in Figure~\ref{fig:overview}, \projectname\ has two main components: the monitoring data encoding and the CNN model. The first component, the monitoring data encoding component, performs a series of operations on the raw monitoring data. It creates job signatures encoding the time-series data and represents the jobs as images. The second component is a CNN model that is customized to learn features from job signatures and to classify these job signatures. \projectname\ operates in two phases. The first phase is the offline training phase, where the CNN model is trained from a labelled job signature dataset. The training phase can be enhanced with transfer learning~\cite{pan2009survey}, where the convolutional layers can be transferred from a trained job detection model and only the layers that make predictions need to be trained. The runtime recognition phase is where \projectname\ operates to detect and to predict the applications by job signatures.

Classification model should not be limited to classify known applications already seen in the training phase. To make \projectname\ practically useful, we introduce confidence thresholds to help identify applications. When the prediction probability exceeds the defined threshold, \projectname\ labels the job with the application name; otherwise, it marks the observation as \emph{unknown}, indicating the job is likely to be a new application.

\subsection{Construction of Job Signature}\label{signature}

\begin{figure*}[t]
     \centering
     \begin{subfigure}[b]{0.35\textwidth}
        \includegraphics[width=0.95\linewidth]{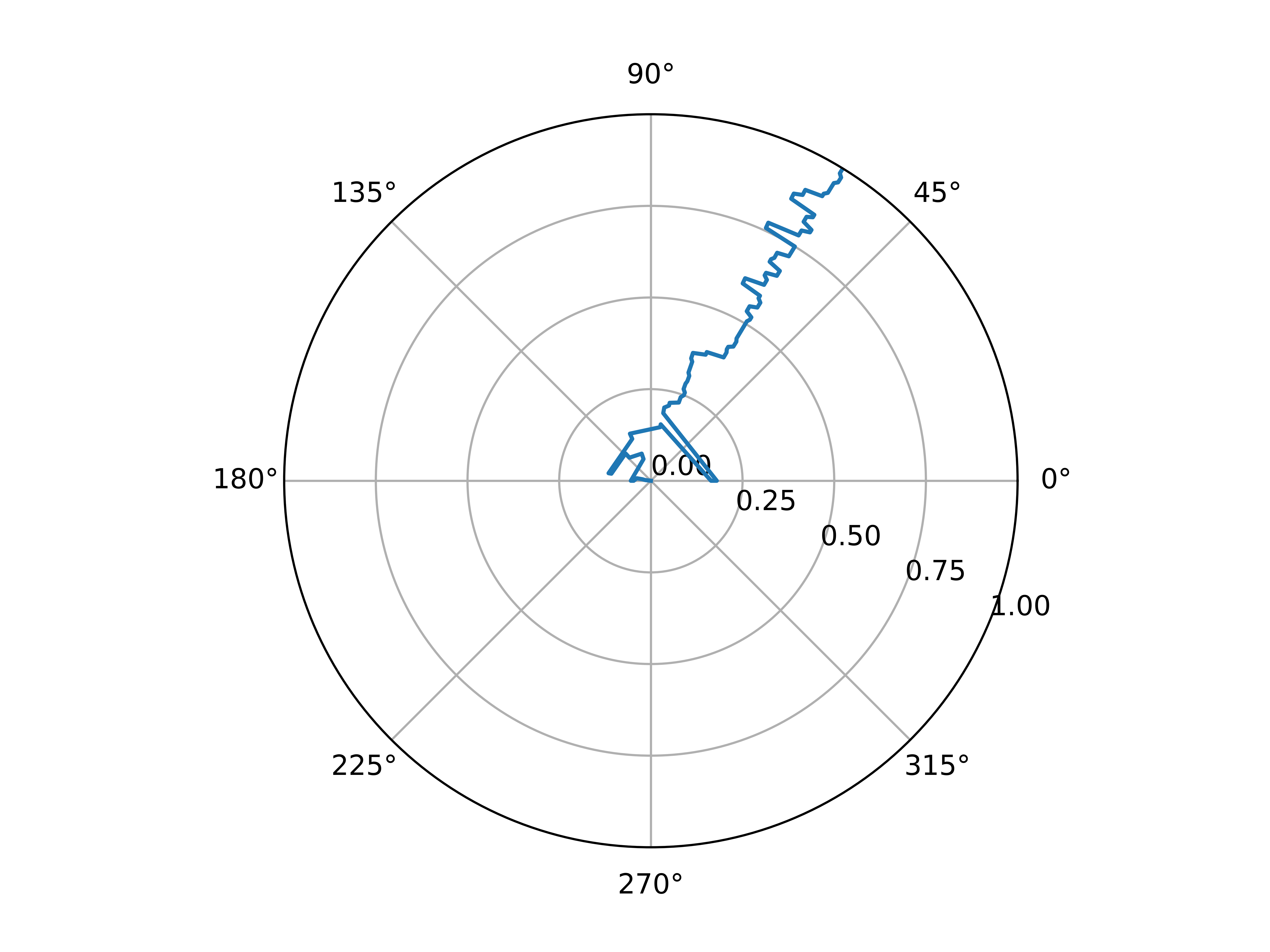}
        \caption{Polar Coordinate of the Normalized Trace}
    \label{fig:polar}
     \end{subfigure}
     \begin{subfigure}[b]{0.50\textwidth}
        \includegraphics[width=0.95\linewidth]{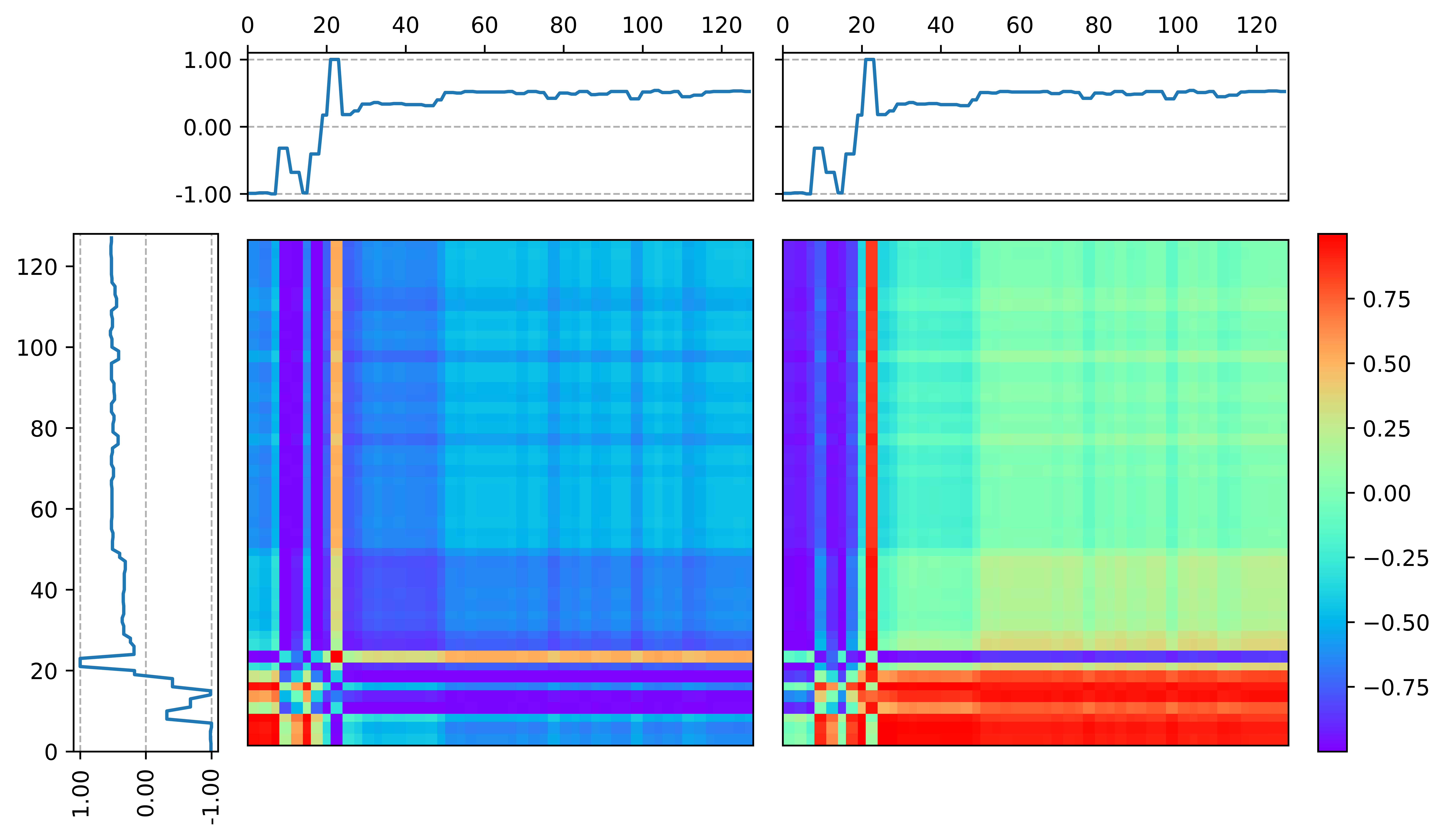}
        \caption{Gramian Angular Summation (Left) and Difference (Right) Fields}
    \label{fig:gramian}
     \end{subfigure}
     
    \caption{Steps of Gramian Angular Field Conversion. Each encoded image has a resolution of $128 \times 128$.}
    \label{fig:gaf}
\end{figure*}

\subsubsection{Resampling Traces}

To construct job signatures that can be measured for the similarity between pairs of signatures, it requires the time series traces to have equal length. Given a trace, we resample it with a predefined length $l$. For a trace $T$ of length $n$,  we create $T'$ by sampling the data points from $T$ so that $n'=l$, where $n'$ is the length of the sampled trace $T'$. Note that the predefined length is usually set to a relatively small value to reduce the compute time when calculating the similarities. 

Considering most jobs on HPC run for hours or even days, their corresponding monitoring traces are usually longer than the predefined length, i.e, $n>l$. In this case, we downsample the traces. For each set of $\lfloor n/l \rfloor$ data points, we apply one of the functions, such as mean, max, min, median, to calculate the aggregated value. Assuming we set the predefined length to be 120, to resample the power consumption trace of a 10-minute job (i.e., the trace has 600 timesteps in total), every 5 data points should be aggregated. Figure~\ref{fig:downsampling_raw} and Figure~\ref{fig:downsampling} illustrate the procedure of resampling the power consumption traces of a job using the \textit{mean} value. The original trace contains more than 14,000 timesteps, while after resampling, the trace length becomes 128 timesteps. 

In case that $n<l$, i.e., the length of a job metric trace is less than the predefined length $l$, we upsample the original traces. Specifically, we add $\lfloor l/n \rfloor$ paddings between consecutive data points and fill with the previous value. Figure~\ref{fig:upsampling_raw} and Figure~\ref{fig:upsampling} show the traces before and after upsampling. After resampling, all traces have the length of $l$, irrespective of the duration of the job. The corresponding resampled traces $T'$ will be used for further processing.

\subsubsection{Converting 1D time series to 2D images}\label{conversion}

As shown in Figure~\ref{fig:resampling}, the monitoring traces and the corresponding resampled traces are univariate time series. To take advantage of the feature learning in deep learning architectures, we convert 1D time series to 2D images. 
% in computer vision to learn features and find similarities in time series data,

We utilize the Gramian Angular Field (GAF) to transform time series into images~\cite{wang2015encoding}. Specifically, the time series data is first normalized or scaled into the range of $[-1, 1]$. The normalized time series data is then represented in a polar coordinate instead of the typical Cartesian coordinate. A Gram Matrix like operation is applied on the resulting angles to construct 2D images.

Given a time series trace $T = \{t_1, t_2, ..., t_n\}$ of $n$ timestamps, we rescale $T$ to have the interval $[-1, 1]$ by the equation below:

\begin{equation}\label{eq:rescale}
\tilde{t_i} = \frac{(t_i - max(T)) + (t_i - min(T))}{max(T) - min(T)}
\end{equation}

The value of the time series and its corresponding timestamp need to be accounted for so that no information is lost. These two quantities are expressed with the angle and the radius in polar coordinates, respectively. Mathematically, the angle is computed by $arccos(\tilde{t_i})$, which lies within $[0, \pi]$, and the radius variable is calculated by $i/n$, which is in $[0, 1]$. The point can be expressed in polar coordinates ($\phi_i$, $r_i$), where:

% Therefore, the point ($\phi_i$, $r_i$) in polar coordinates can be expressed as:

\begin{equation}
\begin{cases} 
    \phi_i = arccos(\tilde{t_i}), & -1\leq \tilde{t_i} \leq 1 \\
    r_i = \frac{i}{n}, & 0\leq i \leq n
\end{cases}  
\end{equation}

The encoding function is a composition of bijective functions, producing one and only one result in the polar coordinate system. In addition, as opposed to Cartesian coordinates, polar coordinates preserve temporal dependency through the $r$ coordinate. An example of a trace represented in polar coordinates is shown in Figure~\ref{fig:polar}, which is transformed from the normalized trace of Figure~\ref{fig:upsampling}.

The temporal correlations between each pair of data points $(t_i, t_ j)$ are computed by considering the trigonometric summation ($cos(\phi_i+\phi_j)$) or subtraction ($cos(\phi_i-\phi_j)$), leading to the Gramian Matrix called \emph{Gramian Angular Summation Field (GASF)} or \emph{Gramian Angular Difference Field (GADF)}, respectively. The GASF is defined as follows:

\begin{equation}
GASF = \begin{bmatrix} 
    cos(\phi_1 + \phi_1) & \dots & cos(\phi_1 + \phi_n)\\
    cos(\phi_2 + \phi_1) & \dots & cos(\phi_2 + \phi_n)\\
    \vdots & \ddots & \vdots\\
    cos(\phi_n + \phi_1) & \dots & cos(\phi_n + \phi_n) 
    \end{bmatrix}
\end{equation}

Through the GASF or GADF conversion, the diagonal $G_{i,i}$ contains the original value of the scaled time series, while $G_{i,j}$ represents the relative correlation by superposition of directions with respect to time interval $|i-j|$. Other details on time series encoding can be found from~\cite{wang2015encoding}. In Figure~\ref{fig:gramian}, we illustrate the encoded 2D images with GASF and GADF for the trace in Figure~\ref{fig:polar}. To keep it concise, we use GASF as the 1D time series conversion algorithm and refer to GASF as GAF in the following discussion.

\subsubsection{Encoding Multiple Time Series Traces}

\begin{figure*}[t]
    \centering
    \includegraphics[width=0.95\linewidth]{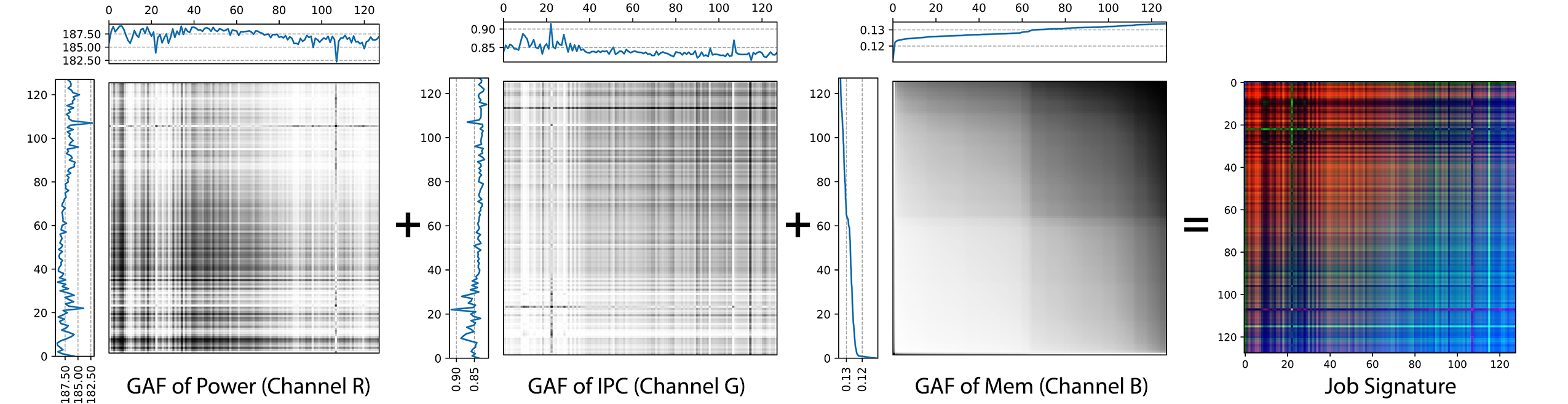}
    \caption{Encoding Multiple Time Series Traces into a Job Signature with a Resolution of $128 \times 128$.}
    \label{fig:encoding}
\end{figure*}

The procedure presented in~\cref{conversion} only converts one monitoring trace to a 2D image. Since we have multiple monitoring traces corresponding to the same job and we do not want to lose the correlation between each pair of traces. The question becomes how we can encode multiple time series to a single image such that a deep learning architecture can understand. To solve this problem, we are inspired by the concept of RGB channels, where each RGB channel emphasizes different aspects of the original image. Similarly, the power consumption trace, IPC trace, and memory used trace can be considered as the R channel, G channel, and B channel of the encoded image, respectively. 

The output of GAF conversion is nothing but a 2D matrix of floating-point numbers that fall in $[-1, 1]$. To visualize the job signature, we rescale the GAFs of the above-mentioned three traces to be in $[0, 255]$ by using the below equation:

\begin{equation}
    \widetilde{GAF} = \lfloor \frac{GAF + 1}{2} * 255 \rfloor
\end{equation}

The $\widetilde{GAF}$ contains the pixel value of the gray scale image. When combining $\widetilde{GAF}_{power}$, $\widetilde{GAF}_{ipc}$, $\widetilde{GAF}_{mem}$ as three channels of RGB, we create a single color image. Figure~\ref{fig:encoding} shows the GAFs of the power, IPC and memory usage of a job and its corresponding encoded job signature. It is worth noting that rescaling GAF to the range $[0, 255]$ is only for the purpose of visualizing the job signature while the original GAF falling in $[-1, 1]$ can be directly fed in the Convelutional Neural Network. 

It is important to note that the channel-like encoding methodology is not limited to \emph{three} channels. Each time series trace is converted to a $l\times l$ matrix by the procedure presented in \cref{conversion}. An encoded job signature of three channels is a $l\times l \times 3$ matrix. Encoding one more metric is simply adding another dimension in the matrix. More formally, a job signature of $c$ metrics is a $l\times l \times c$ matrix. Even though it is not straightforward to be visualized when $c$ is larger than 3, the CNN model can ``understand'' and analyze the high-dimensional matrix.

\subsection{Classification Model}\label{model}
Using the methodology presented in~\cref{signature}, the monitoring traces of jobs can be represented in images (i.e., job signatures). Therefore, detecting and identifying applications of jobs become an image recognition problem. In this subsection, we first introduce the CNN, a deep learning technique that has been widely used in image classification problems and achieved promising results in many domains. Then, we present the CNN architecture that is specifically customized for classifying job signatures.

\subsubsection{Deep learning using CNN}

The performance of conventional machine learning techniques depends heavily on data representation, which requires lots of efforts to design preprocessing pipeline and feature engineering. Such feature engineering is labor intensive and lacks the ability to extract discriminative information from the data~\cite{bengio2013representation}. Deep learning, on the other hand, explores the possibility to feed raw data to the algorithm and automatically discover the features needed for detection or classification. The key concept of deep learning is to transform the representation at a lower level into a representation at a higher and more abstract level; and with composition of such transformations, complex functions can be learned~\cite{lecun2015deep}. As a widely-used deep learning technique, CNNs have been successful in image classification problems.

% The architecture of a CNN consists of multiple stacked feature learning  stages, including convolution, non-linearity activation and pooling operators, followed by a fully-connected (FC) layer for classification~\cite{lecun2015deep}. The input and output of the feature learning stage are sets of arrays called \emph{feature maps}. For a color image input, each feature map is a 2D array containing a color channel. Each feature map of the output represents a particular feature extracted at the locations of the associated input~\cite{lecun2010convolutional}. After a series of feature learning stages, the feature outputs will be flattened into vector and fed to a fully-connected neural network (i.e., the classification layer), in which an activate function such as softmax or sigmoid is used to classify the outputs. 

% A supervised training of CNN is performed by two main steps: propagation and parameter update. The model feeds input and computes layer by layer to generate a final output that is compared with the correct result, on the basis of which the propagation error is calculated. The error is then back-propagated into the CNN, and the model parameters in all layers are updated in a direction of reducing the error. For faster convergence, the stochastic gradient descent (SGD) is used to update the parameters. Propagation and parameter update are repeated until the model reaches a satisfactory validation error. Details about the CNN architecture and training can be found in papers~\cite{lecun1998gradient, lecun2010convolutional}.

\subsubsection{Customized CNN Architecture for Job Signatures}

\begin{figure*}[t]
    \centering
    \includegraphics[width=0.95\linewidth]{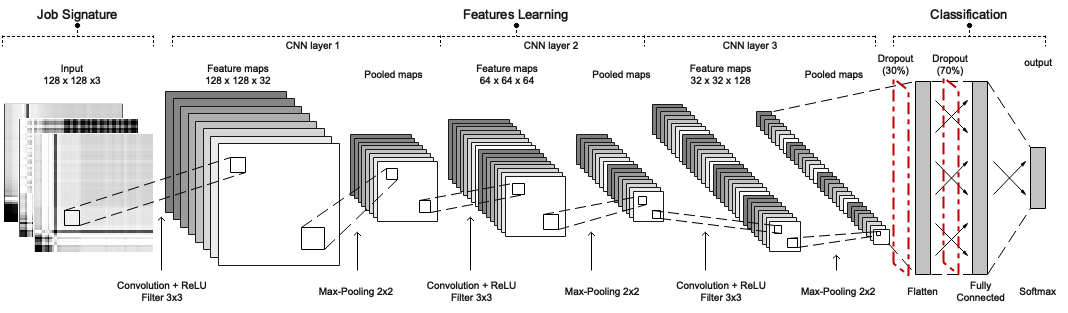}
    \caption{Customized CNN Architecture for Classifying Job Signatures.}
    \label{fig:cnn}
\end{figure*}

The CNN architecture for classifying job signatures is designed as shown in Figure~\ref{fig:cnn}, which contains the following customized layers and parameters:
%  Contribution of our customized CNN model. 1. Input layer for job signatures. 2. Layers for feature extraction of Job signatures. 3. Classification of Job signatures.

\textbf{1. Input layer for job signature}: The inputs are job signatures of encoded multiple time series traces, as presented in~\cref{signature}. The resolution of the job signature is $128 \times 128$ and the number of channels equals to the number of traces encoded in the job signature (i.e., three in our dataset). Note that for other cases that job signatures encode more than three monitoring traces, the input shape should be set accordingly. 
% However, the job signatures cannot be easily visualized in this scenario.
    
\textbf{2. Layers for feature extraction}: In our model, we have \emph{three} layers of CNN to learn features from the job signature, each of which contains one convolution layer to extract features and one pooling layer to reduce the spatial dimension of the convoluted image. The convolution layer extracts image features by convolving the job signature with a set of kernels and produces one feature map for each kernel. We apply the activation function \emph{ReLU} on the convoluted features. It replaces negative values with zero and keeps the positive value. The output of ReLU will be the input for the pooling layer.

The first convolution layer learns 32 kernels, and the second and third convolution layers learn 64 and 128 kernels, respectively. We set the kernels to be size of $3 \times 3$ and use the \emph{same} padding in the convolution to maintain the dimension of output as input. All three pooling layers perform max pooling that returns the maximum value from the portion of the image covered by a window of size of $2 \times 2$.
    
\textbf{3. Layers for classification}: After going through the feature extraction layers, the customized CNN model flattens the final output and feeds it to a regular neural network for classification. The flattened vector is connected to a fully connected layer, through which non-linear combinations of the features can be learned. After passing through the fully connected layer, we use the \emph{softmax} activation function in the last layer to get the probabilities of the input job signature being in a particular class.
    
To prevent the CNN model from overfitting and improve the generalization of the CNN model, we add several dropout layers in the model to randomly disable neurons during training, as shown in the dashed red parallelograms in Figure~\ref{fig:cnn}. In our experiments, the dropout rate before the flatten layer and in the fully-connected layer is set as $30\%$ and $70\%$, respectively.

\section{Experimental Evaluation}\label{experiments}
In this section, we introduce the experimental evaluation of our proposed classification model on a real-world dataset collected from the Cori system. Particularly, we first introduce the dataset built based on the proposed encoding methodology and then, we compare the performance of our model with the baseline methods in terms of accuracy. 
% F1 score (harmonic mean of precision and recall), which is a widely used measurement for evaluating classification models.
% The effects of using dropout layers in the CNN model will also be discussed. 

\subsection{Dataset}

\begin{table*}[ht]
    \centering
    \caption{Dataset of Job Signatures}
    \label{tab:exit}
    \begin{threeparttable}
    \begin{tabular}{c|c|c|c|c}
        \hline
        \rowcolor{Gray}
        \textbf{Application} & \textbf{\# of Jobs} & \textbf{\# of Nodes\tnote{*}} & \textbf{Runtime(s)\tnote{*}} & \textbf{Description} \\
        \hline
        BerkeleyGW & 1899 & 1 / 2048 & 60 / 172867 & For quasiparticle excitations and optical properties of materials. \\
        \hline
        Espresso & 2000 & 1 / 2048 & 61 / 172858 & For electronic-structure calculations and materials modeling at the nanoscale.\\
        \hline
        Gromacs & 1937 & 1 / 64 & 61 / 172836 & For simulations of proteins, lipids and nucleic acids.\\
        \hline
        LAMMPS & 1999 & 1 / 64 & 60 / 172866 & For molecular dynamics  with a focus on materials modeling.\\
        \hline
        NWChem & 1992 & 1 / 384 & 71 / 169209 & For computational chemistry.\\
        \hline
        VASP & 2000 & 1 / 128 & 69 / 172864 & For ``ab-initio'' quantum-mechanical molecular dynamics (MD) simulations. \\
        \hline
        WRF & 1860 & 1 / 256 & 72 / 174608 & for atmospheric research and operational forecasting applications. \\
        \hline
        aims & 2000 & 1 / 192 & 71 / 172959 & For ``ab-initio'' molecular simulations.\\
        \hline
        chroma & 1996 & 2 / 137 & 488 / 147447 & For lattice Quantum Chromodynamics calculations (LQCD).\\
        \hline
        cp2k & 1993 & 1 / 128 & 60 / 174608 & For quantum chemistry and solid state physics.\\
        \hline
        e3sm & 1989 & 1 / 2048 & 60 / 173402 & For earth system modeling, simulation and prediction.\\
        \hline
        su3 & 2000 & 1 / 36 & 65 / 20277 & For lattice Quantum Chromodynamics calculations (LQCD)\\
        \hline
        
    \end{tabular}
    \begin{tablenotes}[para,flushleft]
     {*The values on the left and right side of `/' indicate the minimum and maximum values of the corresponding characteristics of the jobs.}
    \end{tablenotes}
    \end{threeparttable}
    \label{table:dataset}
\end{table*}

To the best of our knowledge, there are no publicly released monitoring traces with labelled application information collected from production HPC systems. Ates el al.~\cite{ates2018artifact} published a dataset containing the monitoring data of benchmarks and proxy applications. Google published its cluster traces that do not have application information~\cite{clusterdata:Wilkes2020a}. Therefore, in order to train the CNN model for detecting applications running in production HPC systems, we built our own dataset.

On Cori system, some other research groups had implemented the job metadata management service, where the metadata of jobs running on Cori are stored and managed in a MySQL database. The job metadata has a field named `application\_name' which is derived from the job submission script. However, based on our analysis, about $42.4\%$ of derived names do not reflect the application accurately. Many of them have names like `test', `bugs', `b.sh', etc. Therefore, to build a reliable labelled dataset, we select jobs that have accurate application names through the job metadata service. In addition, since Cori system has the KNL partition and the Haswell partition, the monitoring metrics of the same application (even with same configurations and same input files) running on different architectures could potentially have large variations. To avoid discrepancy caused by the architecture, we only focus on the jobs running on the same architecture and we select KNL jobs as our dataset. Besides, we discard short jobs that run for less than 60 seconds since they are likely for testing purposes. 

The job IDs of the selected jobs are used with NERSC\_LDMS~\cite{nerscldms} to obtain the corresponding monitoring traces. Considering that a job may have multiple steps and that job steps may correspond to different applications, we treat the job steps of a job separately. In addition, since a job may use multiple nodes (nodes are exclusively used by the job on Cori), we aggregate the monitoring metrics from all involved nodes and use the average time-series traces to build the job signature. 

We select 23,665 jobs from 12 different applications to build the job signature dataset after eliminating those jobs that do not have monitoring metrics (due to data collection errors or historical traces are purged) and balancing the number of jobs of each application as much as possible. Details are listed in Table~\ref{table:dataset}. The numbers of allocated nodes for a job vary from 1 to a maximum of 2,048. The \emph{BerkeleyGW, Espresso} and \emph{s3sm} have extremely large-scale jobs where all cores of 2,048 nodes are used, corresponding to 131,072 physical cores. The selected jobs also cover a large range of runtime variations. The shortest job, as defined earlier, runs for only 60 seconds. The longest job, limited by the Quality of Service of the KNL partition, runs for 48 hours. The dataset can be found in a separate submission of artifacts.

It is worth mentioning that, it is up to users to set the resamling length $l$ and define the resolution based on their computing capability. Nevertheless, the CNN model for images of larger resolution usually requires more complex architecture and the time for fine tuning and training the parameters of the architecture becomes high. In our experiments, we set the resampling length to be 128 and use a resolution of $128 \times 128$ for job signatures.

\subsection{Baseline Methods}
To compare the performance of our proposed CNN model with state-of-the-art methods for detecting HPC jobs, we examine several other performance metrics based approaches. The Power Signature~\cite{combs2014power} and Taxonomist~\cite{ates2018taxonomist} are two representative methods that use statistic features of the time series monitoring data to build the classification model. The statistics include the \emph{minimum}, \emph{maximum}, \emph{mean}, \emph{standard deviation}, \emph{skew}, \emph{kurtosis} and the $5^{th}$, $25^{th}$, $50^{th}$, $75^{th}$, $95^{th}$ \emph{percentiles}. Using the same monitoring metrics as \projectname, namely power consumption, IPC and memory usage of the selected jobs, we extract statistical features and build several classifiers as baselines. These classifiers are Random Forest (RF), linear Support Vector Classifier (Liner-SVC), and non-linear Support Vector Classifier (SVC). In addition, we include a Random Forest model that uses only statistical features extracted from the power consumption trace. All of these models use default hyperparameters.

\subsection{Experiment Setup}

\begin{figure}[t]
    \centering
    \includegraphics[width=0.97\linewidth]{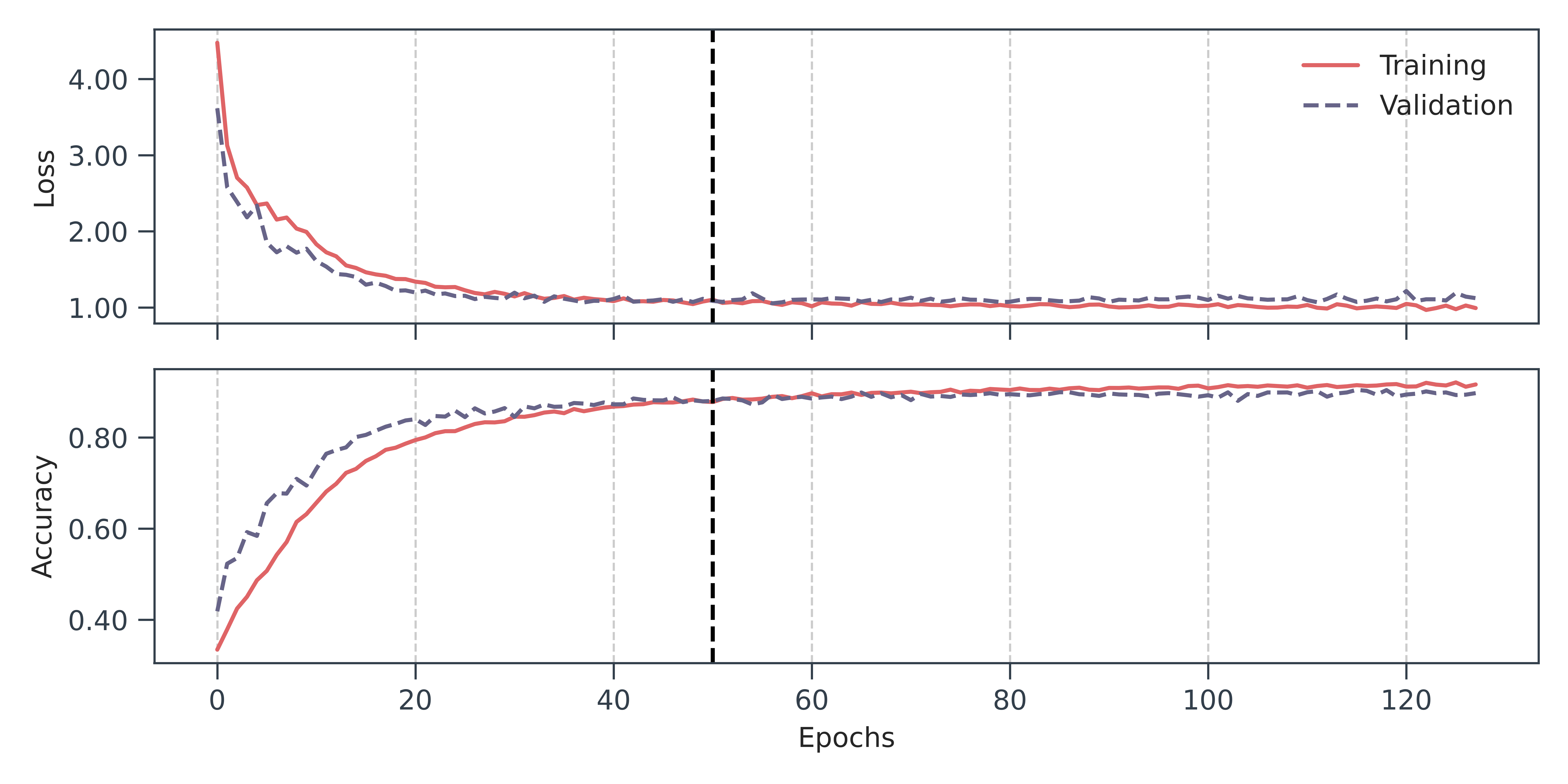}
    \caption{Loss (top) and accuracy (bottom) of the training and validation data. The vertical dashed line indicates the determined epoch of 50.}
    \label{fig:epochs}
\end{figure}

The \projectname\ and baseline models are trained and evaluated on a Cori GPU node, where 2 Skylake CPUs, 8 NVIDIA Tesla V100 GPUs and 384 GB DDR4 memory are provided. The proposed CNN is implemented using the \emph{TensorFlow Keras} library in a Python 3.9 environment. The baseline models are implemented in Python leveraging \emph{scikit-learn} library.

% The dataset for the baseline methods are built upon the statistic features of the job traces that are used exactly for generating the job signatures. 

We divide the dataset into 60\% for training, 20\% for validation, and 20\% for testing. To mitigate overfitting and to increase the generalization of \projectname\ model, we determine the optimal epoch number by examining the loss and accuracy trends on training and validation data. As shown in Figure~\ref{fig:epochs}, after epoch reaches to 50, the training loss gets lower than validation loss  while the accuracy of validation does not improve. Therefore, we set epoch to 50. While evaluating the classification performance on each application, we use ten-fold stratified cross validation to divide the dataset into ten disjoint partitions and evaluate model performance with each partition.

We test the classification performance on different confidence thresholds. For each baseline classifier, we use the one-vs.-rest version of that classifier such that it produces a set of real-valued prediction scores for its decision instead of a labeled class. In \projectname\ model, the softmax function assigns probabilities of classes in each prediction. We compare the prediction probabilities with the confidence thresholds to determine the predicted class. For the prediction score $s_i$ of class $ c_i, i\in \{1,...,k\} $, the classifier assigns the label of the class by the following expression:

\begin{equation}
\begin{cases} 
    c_i, where\ s_i = max(s_1, ..., s_k), & \text{if  $s_i$ $\geq$ threshold} \\
    unknown, & \text{otherwise}
\end{cases} 
\end{equation}

In other words, when the maximum value of the prediction scores is larger than the threshold, the class with the largest score is the predicted class. Otherwise, the classifier labels the observation with \emph{unknown}. When the confidence threshold is 0, it is the same with the vanilla multi-class classifier. 

\subsection{Experimental Results}

We evaluate the capability of \projectname\ in identifying applications in this subsection. First, we evaluate the classification performance on different confidence thresholds and examine the performance on each application. Then, we assess the performance of identifying applications that have not been trained with the models. In addition, since the job signatures encode temporal information of the monitoring metrics, we use a subset of the monitoring data to construct partial signatures to evaluate the \projectname\ model.

\subsubsection{Classification Performance}

\begin{figure}[t]
    \centering
    \includegraphics[width=0.95\linewidth]{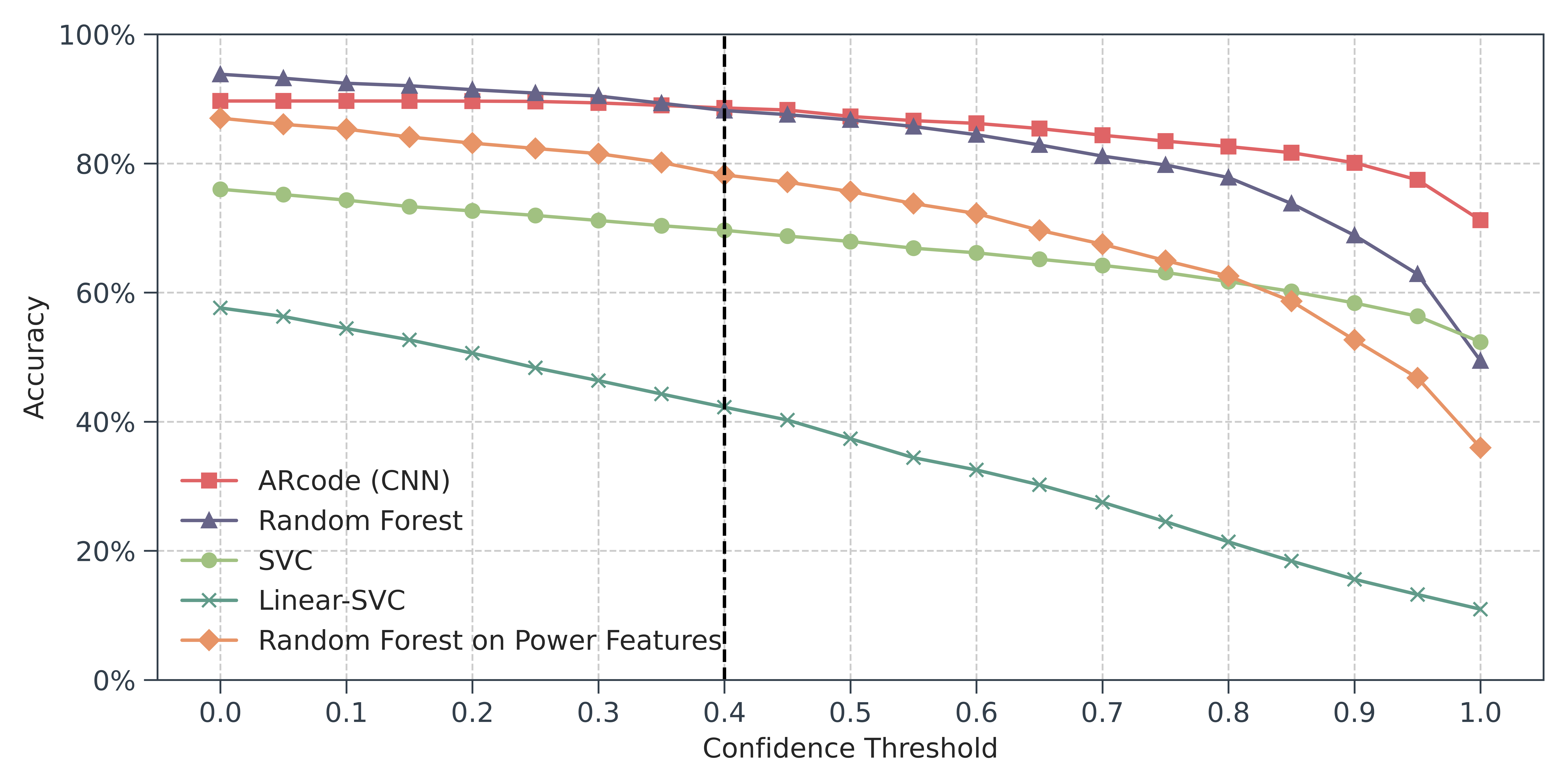}
    \caption{Accuracy of \projectname\ and baseline classifiers at different confidence thresholds. The vertical dashed line indicates the confidence threshold (0.4) above which \projectname\ outperforms all other classifiers.}
    \label{fig:accuracy_cmp}
\end{figure}

\begin{figure}[t]
    \centering
    \includegraphics[width=0.95\linewidth]{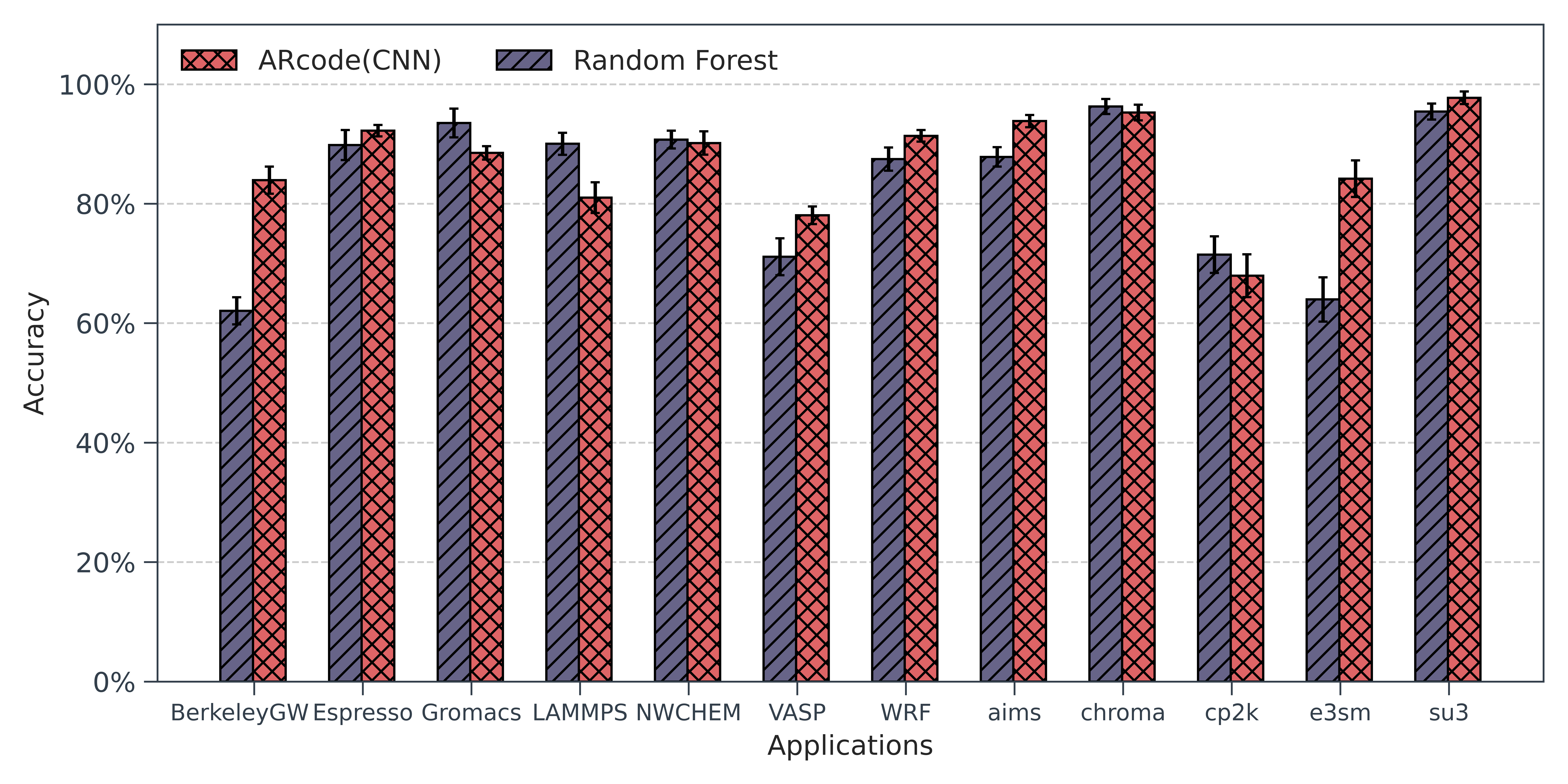}
    \caption{Accuracy of classifiers on each application at a confidence threshold of 0.8.}
    \label{fig:accuracy_app}
\end{figure}

Figure~\ref{fig:accuracy_cmp} depicts the accuracy of \projectname\ and baseline models on different confidence thresholds. As we can see from the figure, the Linear-SVC model has the worst performance at all confidence thresholds. When the confidence threshold is below 0.4, Random Forest outperforms all other classifiers, with the highest accuracy of 93.81\% at a confidence threshold of 0. \projectname\ follows closely with an accuracy of 89.67\%. When the confidence threshold is above 0.4, the accuracy of Random Forest decreases and \projectname\ has the best performance. At a threshold of 1.0, the SVC model performs a little better than Random Forest, but it is still 18.87\% worse than \projectname. At a confidence threshold of 0, the accuracy of Random Forest on power features is close to \projectname\ , but it decreases significantly as the confidence threshold increases. In summary, the accuracy of all classifier decreases with increasing confidence thresholds, but the slowest decrease rate is observed for \projectname. 

We further examine the classification performance of \projectname\ and Random Forest on each application, as shown in Figure~\ref{fig:accuracy_app}, where we use a confidence threshold of 0.8 in the experiment. From this figure, we have the following observations. First, the accuracy of \projectname\ and Random Forest are very close in most cases. When detecting Gromacs, LAMMPS and cp2k, Random Forest performs slightly better. Second, for some applications such as BerkeleyGW and e3sm, \projectname\ achieves a significantly better accuracy. In BerkeleyGW classification, the accuracy of Random Forest is 62.06\% while \projectname\ achieves 83.94\% accuracy. \projectname\ is also 20.21\% better than Random Forest in detecting e3sm.

\subsubsection{Classification on Novel Applications}

\begin{figure}[t]
    \centering
    \includegraphics[width=0.95\linewidth]{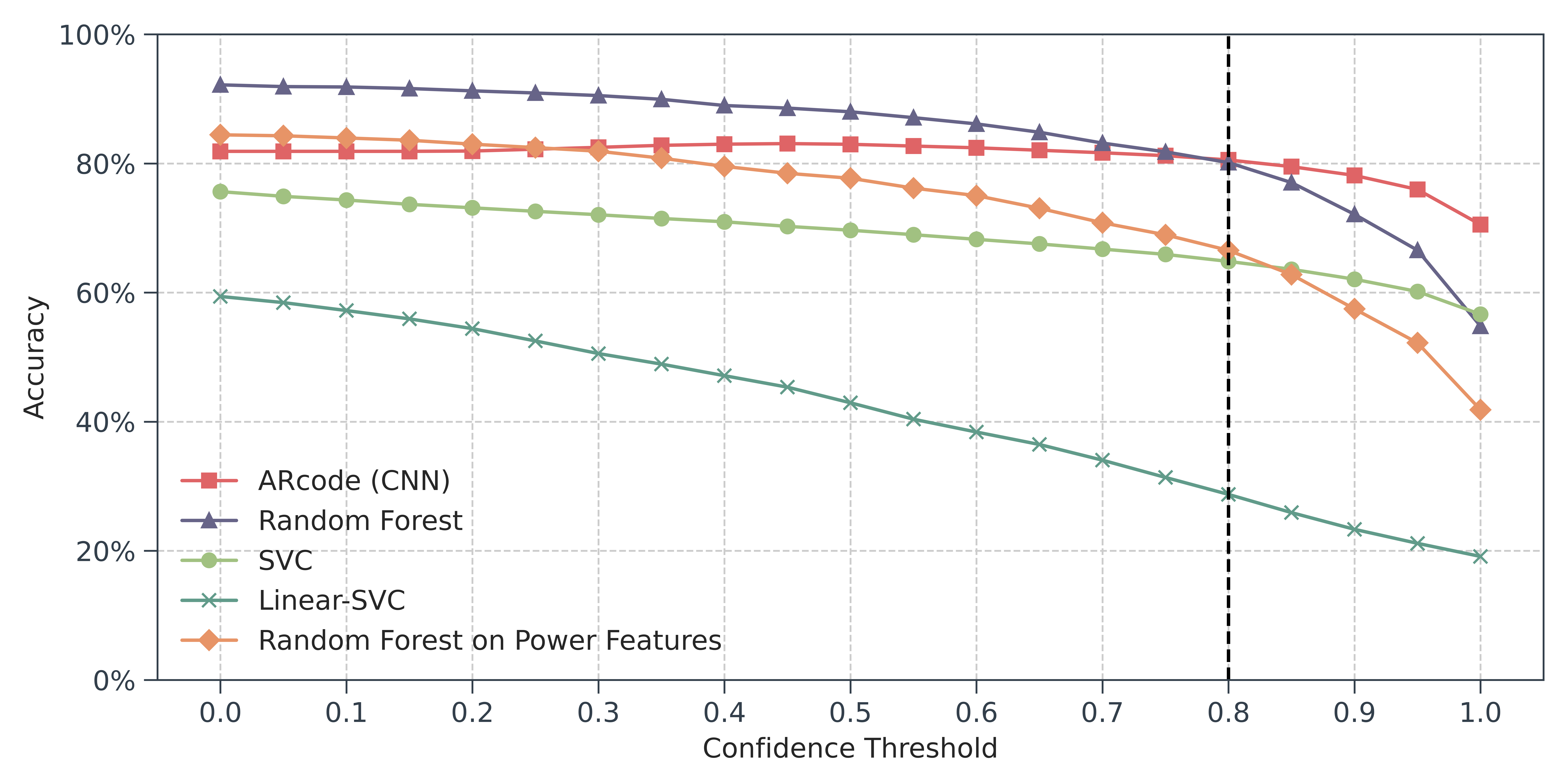}
    \caption{Accuracy of \projectname\ and baseline classifiers on detecting novel applications at different confidence thresholds. \projectname\ achieved the highest accuracy when the confidence threshold is greater than 0.8.}
    \label{fig:accuracy_unknown}
\end{figure}

Considering that in HPC environments, `novel' applications are more prevalent than those trained with the classifier, it is appropriate to define an `unknown' class for all these novel applications. To be practically useful, the classification system must classify both known and unknown novel applications. We evaluate this capability by removing one application from the training set while keeping the testing set untouched. If the removed application in the testing set is predicted to be unknown, we mark it as a correct prediction. 

The classification performance for novel applications is depicted in Figure~\ref{fig:accuracy_unknown}. From this figure, we can observe that Random Forest has the best prediction accuracy when the confidence threshold is below 0.8, while \projectname\ beats Random Forest once the confidence threshold is greater than 0.8. \projectname, similar to its performance in known application detection, has relatively stable prediction accuracy across all confidence thresholds. However, the accuracy of Random Forest drops significantly in large confidence thresholds. 

\subsubsection{Classification using Single Channel}

\begin{figure}[t]
    \centering
    \includegraphics[width=0.95\linewidth]{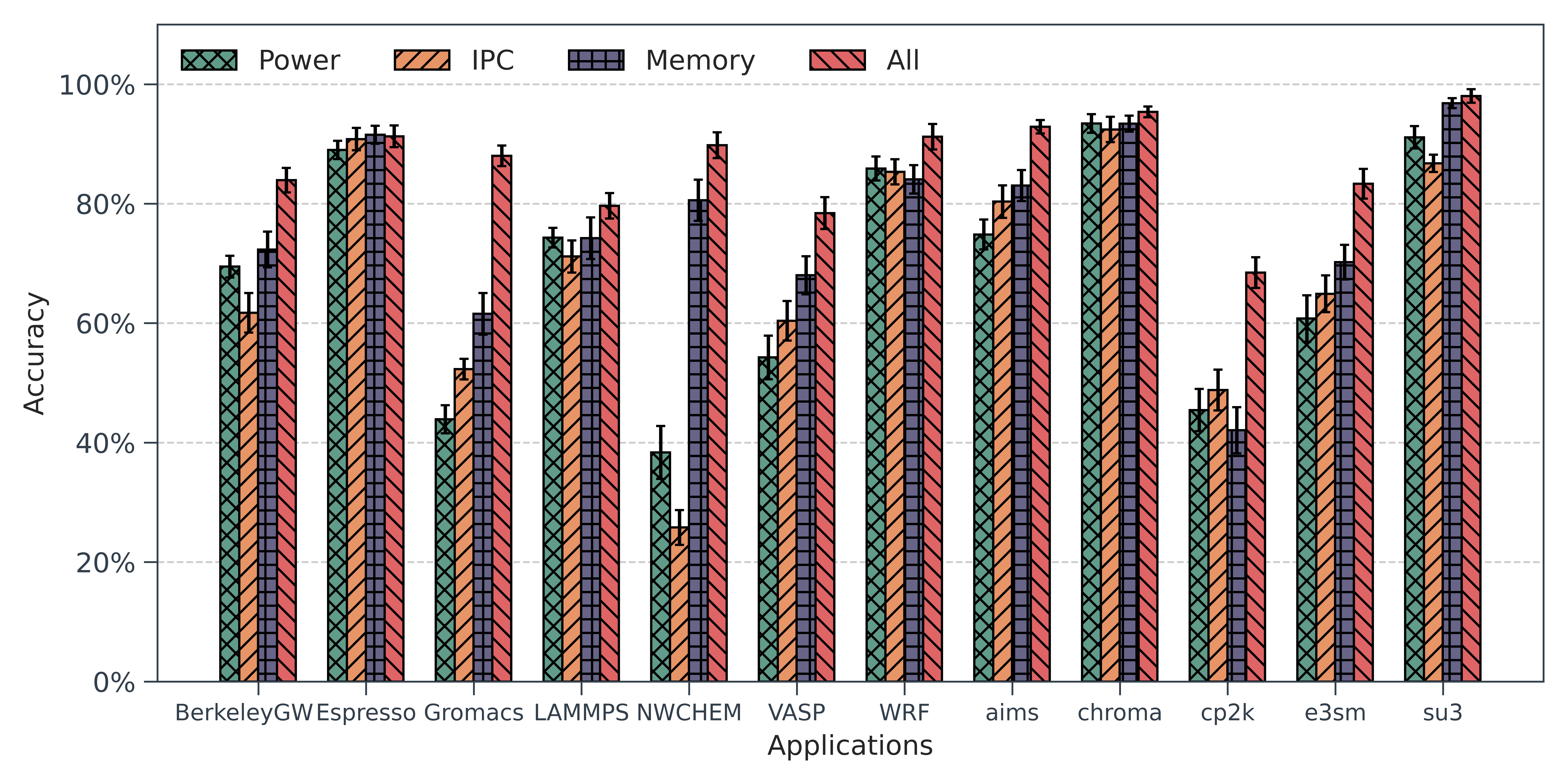}
    \caption{Accuracy of \projectname\ on each application using different channels of the job signature.}
    \label{fig:accuracy_channel}
\end{figure}

In our experiment, \projectname\ encodes \emph{three} channels of monitoring data in the job signature. It is worth knowing whether each channel plays the same importance in classification and whether we can use one channel for detection while still achieving high accuracy. To answer these questions, we train CNN models with each of these channels individually and analyze their classification performance. The results are shown in Figure~\ref{fig:accuracy_channel}. 

From the figure we can see that using any of the channels gives a competitive accuracy when detecting Espresso, LAMMPS, WRF, and chroma compared to using all channels. For NWCHEM, the classification accuracy using the memory channel is significantly better than using the power or IPC channel. We can also see that all channels are important for detecting both Gromacs and cp2k with an accuracy improvement of 26.36\% and 19.66\%, respectively, compared to using only one of the channels. The memory channel is the most representative of these channels. In most applications, using the memory channel in detection has the closest accuracy to using all channels.

\subsubsection{Classification using Partial Signatures}

\begin{figure}[t]
    \centering
    \includegraphics[width=0.95\linewidth]{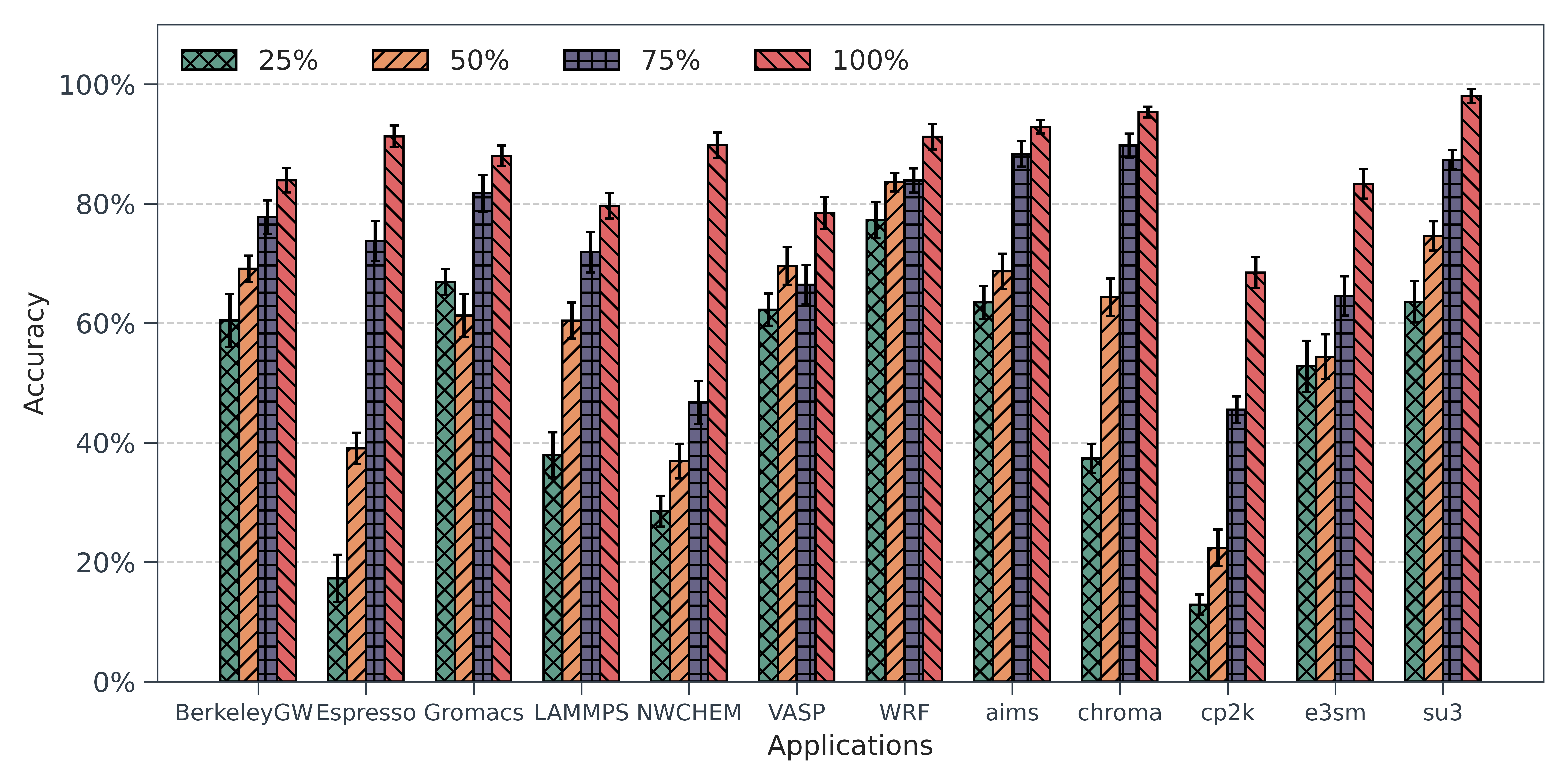}
    \caption{Accuracy of \projectname\ on each application using partial job signature. The partial job signatures are all built from the metrics collected since the startup of jobs.}
    \label{fig:accuracy_partial}
\end{figure}

The job signature retains the temporal information of the time-series monitoring data, and the job signature generated before the end of the job still holds some of the attributes of the full job signature. We use the first 25\%, 50\% and 75\% of the time-series data to create partial job signatures and compare the detection performance with the full job signature (i.e, a job signature built from 100\% of the monitoring data). This experiment is performed to evaluate the usability of the \projectname\ model for detecting running applications, which is not available in the statistics-based models.

The results are shown in Figure~\ref{fig:accuracy_partial}. It is not surprising that the larger the percentage of encoded data, the better the detection performance achieved by \projectname. When encoded with 25\% of the monitoring data, \projectname\ achieves a relatively high accuracy of 77.27\% in detecting WRF, but only 17.27\% and 12.88\% in detecting cp2k and Espresso, respectively. When detecting NWCHEM, the full job signature brings 89.79\% accuracy; however, the 75\% partial job signature gives only 46.73\% accuracy. The improvement from encoding more data varies from application to application. For example, encoding 25\% more data brings an average improvement of 24.67\% for Espresso, but only 4.65\% for WRF.

\subsubsection{Sensitivity to the resampling length}
The performance evaluation described above is based on job signatures with a resolution of $128 \times 128$, i.e., the performance metrics are resampled to a length of 128. It is valuable to have knowledge of the sensitivity of the classification accuracy in terms of the resampling length such that an appropriate resolution can be chosen to achieve a balance between accuracy and training overhead. To explore this, we further build two job signature datasets with resampling lengths of 32 and 64, respectively, and use the same CNN architecture to train these models.

\begin{figure}[t]
    \centering
    \includegraphics[width=0.95\linewidth]{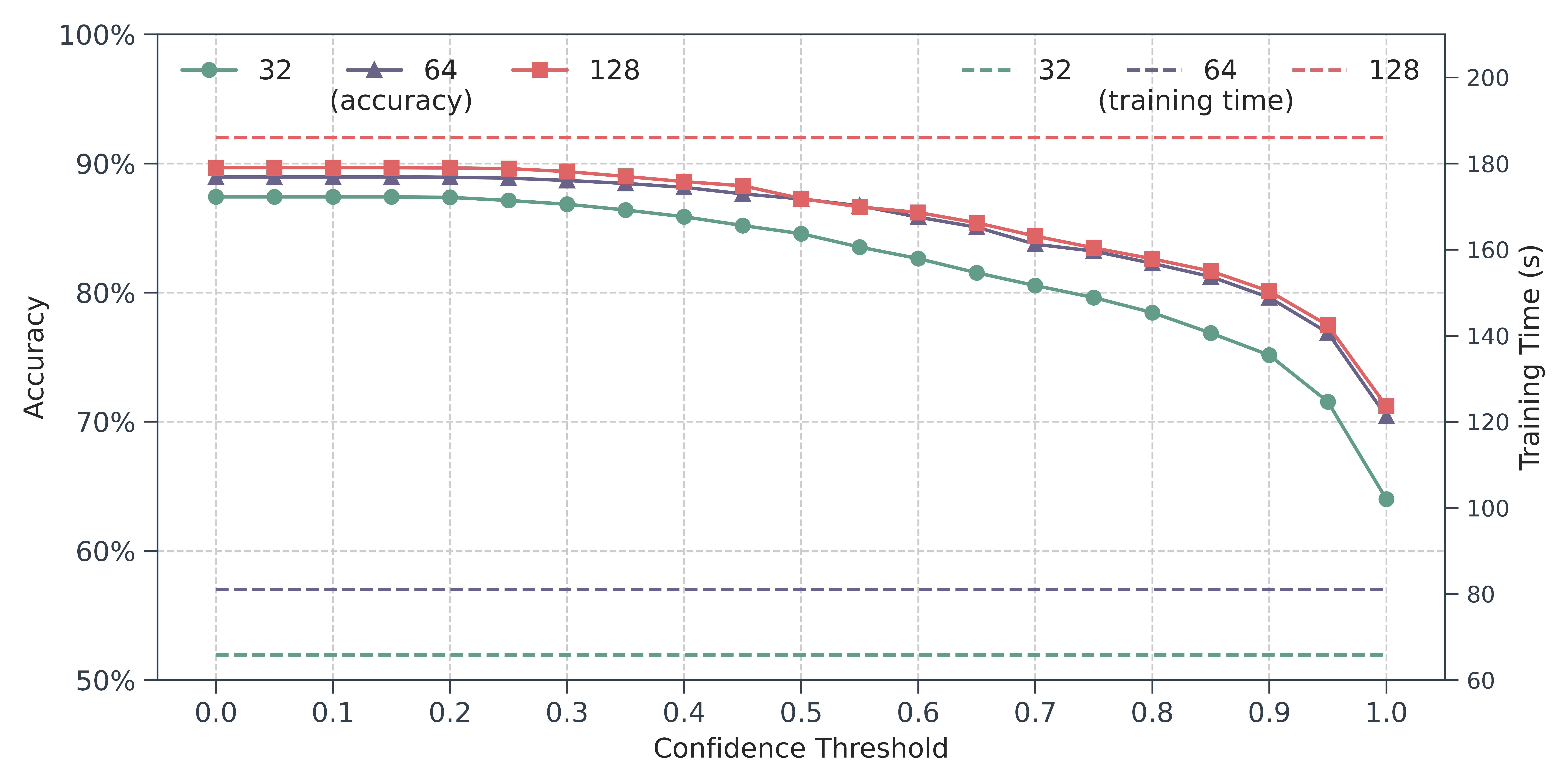}
    \caption{Accuracy and training time of \projectname\ using different resampling lengths. The horizontal dashed lines indicate the training time.}
    \label{fig:accuracy_resolution}
\end{figure}

Figure~\ref{fig:accuracy_resolution} depicts the accuracy (solid lines) and training time (horizontal dashed lines) of \projectname\ using different resampling lengths. Note that the training time is constant across different thresholds for the same resampling length, as illustrated by horizontal dashed lines. This is because training time does not vary according to confidence thresholds, which are set during the prediction of applications. As illustrated from the figure, the training times vary significantly among these three resampling lengths. It is over 185 seconds for the resampling length of 128. For signatures with resampling lengths of 64 and 32, it is 80 and 65 seconds, respectively. Additionally, we can observe from the figure that the higher resolution it is, the higher prediction accuracy. While the accuracy improvement from the resampling length of 64 to 128 is not significant (only by 0.5\% on average), the accuracy improvement from the resampling length of 32 to 64 is 3.0\% on average.

This result suggests that 64 is the optimal resampling length for building job signatures in our experiment, considering the trade-off between accuracy and training time. Its prediction accuracy is close to that of the highest resolution signature, but only uses 43.2\% of the training time of the latter. Furthermore, the data size of the job signature generated with a resolution of $64 \times 64$ is only 25\% of that with $128 \times 128$. 

% For users that the training overhead is not a concern, the high resolution can be set to achieve high prediction accuracy.

% It is also easy to calculate that the size of the signature of $64 \times 64$ is 25\% of that 

\section{Related Works}\label{relatedworks}

Significant amount of work has been reported in the literature on detecting and classifying applications. Identifying similarities and differences among binary executables was explored in~\cite{flake2004structural, dullien2005graph, gao2008binhunt, bourquin2013binslayer}. These approaches are limited by the fact that, they cannot differentiate the same source program compiled by different compiler toolchains or optimization levels. To overcome this limitation, Blanket Execution~\cite{egele2014blanket} presented a binary differencing algorithm that compares the side effects of functions during executation, which is based on the insight that similar codes have semantically similar execution behavior. 

However, the binary based approach cannot be applied in the HPC environment, where it is impractical to conduct binary differencing among hundreds of thousands of executables. In addition, obtaining and managing binaries of users' applications are not always possible for HPC researchers. Instead of using binaries, Yamamoto et al.~\cite{yamamoto2018classifying} used job scripts as job information to classify applications with text classification techniques. This approach, however, also requires dedicated collection and management infrastructures, which are not as prevalent as monitoring infrastructures for performance metric.

Another line of work aims at characterizing HPC applications through system logs. Liu et al. extracted features from combined logs of multiple subsystems to represent applications and build a machine learning model based on the eXtreme Gradient Boosting (XGB) algorithm to identify HPC applications~\cite{liu2020characterization}. DeMasi et al. collected and extracted features from Integrated Performance Monitoring (IPM) performance logs to fingerprint HPC codes~\cite{demasi2013identifying}. Log analysis is more common in characterizing subsystem and user behavior, related works can be found in~\cite{chunduri2018characterization, lim2017scientific, lockwood2018year, patel2019revisiting}.

Monitoring data based application detection has been explored in~\cite{zou2019fingerprinting, ramos2019accurate, ates2018taxonomist, jakobsche2021execution, combs2014power}.  As a early quantitative study of power consumption of HPC workloads, Combs et al.~\cite{combs2014power} studied the applicability of classifying applications through power consumption traces. Ramos et al.~\cite{ramos2019accurate} relied on performance counters to model, fingerprint and clustering applications. Zou et al.~\cite{zou2019fingerprinting} explored detecting illicit applications in GPU-accelerated HPC workloads. They used performance counters, data movement behavior and resources utilization traces to train machine learning models. Taxonomist~\cite{ates2018taxonomist}, proposed by Ates et al., used over 700 system metrics and a time window spanning the whole execution to extract statistical features. They enhanced the classification model such that unknown applications can be detected. EFD~\cite{jakobsche2021execution} created key-values pairs that link execution fingerprints of system metrics to application and input size information to implement application recognition. Except for using measurements of system metrics, they also used the metrics name, node ID and time interval to create fingerprints. 

These studies, however, are based on the datasets built from the monitoring data of benchmarks and proxy applications, where a relative small range of configurations and input size are shown in the applications. On the contrary, the model and experiment results of \projectname\ are developed and tested on datasets collected from real applications in a large-scale production system. In addition, the models of these studies rely on the manually defined features extracted from time series monitoring data and some studies utilize the input information to enhance the detection model. In this work, \projectname\ alleviates the effort of features engineering and takes advantages of the capability of features learning in CNN model to detect and classify encoded monitoring data. Moreover, \projectname\ retains the temporal information of time-series performance metrics, enabling detecting applications before jobs are finished. This capability bridges a major gap in related works. 
\section{Conclusion}

Existing application detection methodologies on HPC systems either relies on the data that are not prevalent or require intensive effort of feature engineering to build high accuracy models. In this study, we aim to provide a solution that can be easily used and adopted by any HPC site. We have introduced \projectname, an application recognition framework which is effective and extensible with the following characteristics: 1) \projectname\ uses three most common monitoring metrics (i.e., the power consumption of the compute node, instruction per cycle, and memory usage) available through the monitoring infrastructure on diverse HPC architectures. 2) \projectname\ alleviates the effort of feature engineering by leveraging the feature learning capability of CNN models. 3) \projectname's channel-like encoding method allows easy encoding of additional metrics in job signatures. 4) \projectname\ encodes job monitoring data into images, which translates the application recognition problem into an image classification problem. Unlike the statistics-based approaches where the temporal information of monitoring data are lost, the job signature encodes metric variations over the runtime. Therefore, the job signature generated from part of monitoring data can still be used in detection, but with the sacrifice of accuracy.

Although our evaluation is performed on job signatures generated from three common monitoring metrics, HPC researchers and system administrators can select other representative metrics and build job signatures to detect and classify applications with specific characteristics such as CPU intensive applications and I/O intensive applications. In addition, we have seen lots of successful cases of applying image recognition in the medical and automobile industries. Their experience in training high-accuracy models can be used in our model to further improve the recognition accuracy. Moreover, encoding monitoring data in job signatures offers a new perspective of exploring and analyzing the performance monitoring data. In our future work, we will further explore the use of job signatures to predict resource usage, detect anomalies, and identify malicious applications. 

% \pagebreak

% With the job signatures, 

% More importantly, our study offers a new possibility of utilizing monitoring data and 

\balance
\bibliographystyle{IEEEtran}
\bibliography{ref}

\end{document}